\documentclass[apj]{emulateapj}
\usepackage{natbib}
\usepackage{physics}
\usepackage{hyperref}
\hypersetup{
    colorlinks=true,
    linkcolor=blue,
    citecolor=blue,
    filecolor=blue,
    urlcolor=blue
}
\usepackage{amsmath,amssymb}
\usepackage{mathptmx}
\usepackage{mathtools}
\usepackage{euscript}
\usepackage{graphicx}
\usepackage{color}
\newcommand\cii{\ifmmode {\rm [CII]} \else [C{\small II}] \fi}
\newcommand\oi{\ifmmode {\rm [OI]} \else [O{\small I}] \fi}

\defcitealias{Sternberg2014}{S14}
\defcitealias{Bialy2016a}{BS16}
\defcitealias{Wolfire2003}{WMHT}

\begin{document}
\title{Thermal Phases of the Neutral Atomic Interstellar Medium \\
from Solar Metallicity to Primordial Gas}
\author{Shmuel Bialy\altaffilmark{1}$^\star$ \& Amiel Sternberg\altaffilmark{2} \altaffilmark{3} \altaffilmark{4}}
\altaffiltext{1}{Harvard-Smithsonian Center for Astrophysics, 60 Garden street, Cambridge, MA, USA}
\altaffiltext{2}{School of Physics \& Astronomy, Tel Aviv University, Ramat Aviv 69978, Israel}
\altaffiltext{3}{Center for Computational Astrophysics, Flatiron Institute, 162 5th Ave., New York, NY, 10010, USA}
\altaffiltext{4}{Max-Planck-Institut f\"ur extraterrestrische Physik (MPE), Giessenbachstr., 85748 Garching, FRG}
\email{$^\star$sbialy@cfa.harvard.edu}
\shorttitle{Thermal Phases -- from Solar Metallicity to Primordial Gas}
\shortauthors{Bialy \& Sternberg}

\begin{abstract}
We study the thermal structure of the neutral atomic (H {\small I}) interstellar medium across a wide range of metallicities, from supersolar down to vanishing metallicity, and for varying UV intensities and cosmic-ray ionization rates.
We calculate self-consistently the gas temperature and species abundances (with a special focus on the residual H$_2$), assuming thermal and chemical steady-state.
For solar metallicity, $Z' \equiv 1$, we recover the known result that there exists a pressure range over which the gas is multiphased, with the warm ($\sim 10^4$ K, WNM) and cold ($\sim 100$ K, CNM) phases coexisting at the same pressure.
At a metallicity $Z' \approx 0.1$, the CNM is colder (compared to $Z'=1$) due to the reduced efficiency of photoelectric heating.
For $Z' \lesssim 0.1$, cosmic-ray ionization becomes the dominant  heating mechanism and the WNM-to-CNM transition shifts to ever increasing pressure/density as the metallicity is reduced. 
For metallicities $Z' \lesssim 0.01$, H$_2$ cooling becomes important,  lowering the temperature of the WNM (down to $\approx 600$ K), and smoothing out the multiphase phenomenon.
At vanishing metallicities, H$_2$ heating becomes effective and the multiphase phenomenon disappears entirely.
We derive analytic expressions for the critical densities for the warm-to-cold phase transition in the different regimes, and the critical metallicities for H$_2$ cooling and heating.
We discuss potential implications on the star-formation rates of galaxies and self-regulation
theories.

\end{abstract}


\section{Introduction}
\label{sec: intro}

Emission and absorption line studies of the 21 cm hyperfine spin-flip transition of atomic hydrogen (H{\small I}) show that cold ($T<200$ K) dense clumps are commonly embedded within extended warm ($T\sim 8000$ K) diffuse atomic clouds.
The cold neutral medium (CNM) clumps are observed as narrow absorption line features embedded within broad emission line envelopes tracing the warm neutral medium (WNM) component
(e.g., in Galactic observations: \citealt{Dickey1978, Kulkarni1987, Colgan1988, Garwood1989, Kolpak2002, Dickey2003, Heiles2003, Roy2013, Stanimirovic2014a, Patra2018, Murray2018}, 
and in external galaxies:
\citealt{Braun1992, Dickey1993, Mebold1997, Dickey2000, Marx-Zimmer2000, Warren2012, Bolatto2013, Herrera-Camus2017}).
  Remarkably, for the characteristic thermal pressures prevailing in the interstellar gas, theory predicts that a multiphase medium is possible in which the CNM and WNM can coexist \citep[][]{Field1969, Wolfire1995, Liszt2002, Wolfire2003}.
The multiphase nature of the H {\small I} gas is also clearly observed in numerous hydrodynamical simulations of the interstellar medium (ISM)  \citep[e.g.,][]{Hu2016, Valdivia2015, Richings2016, Hill2018}.

In the standard theory for the thermal structures of interstellar H{\small I}, the WNM is cooled by electron impact Ly$\alpha$ emission \citep{Gould1970, Dalgarno1972}, and the cold medium by metal line fine structure cooling, mainly neutral impact \cii 158 $\mu$m emission \citep{Dalgarno1964}.  For a given heating rate and at sufficiently low densities metal line cooling becomes very inefficient and the gas heats up to temperatures ($\sim 8000$ K) at which the Ly$\alpha$ emission limits further heating.  At high densities the metal line cooling becomes effective and the temperature drops to values characteristic of the fine-structure energy level spacings ($\sim 100$ K). Because of the particular shape of the combined Ly$\alpha$ plus fine-structure cooling function (we discuss this in detail), an intermediate range of densities and associated pressures exists at which the gas can be either warm or cold.

Hydrodynamical simulations show that CNM forms at sites where flows of WNM collide and compress the gas \citep[e.g., ][]{Hennebelle1999, Kritsuk2002, Audit2004, Vazquez-Semadeni2006, Gazol2013, Saury2014, Gazol2016a}.
The formation of the dense CNM phase and the multiphase structure of the ISM may play a key role in controlling the star-formation properties of galaxies. 
For example, 
\citet{Elmegreen1994, Schaye2004} have argued that the 
 low pressure prevailing at the outskirts of galaxy discs do not allow the formation of the cold phase resulting in a shut-off of star-formation.
The multiphase phenomena where CNM and WNM phases coexist at the same pressure may provide a feedback loop for star-formation in galaxies
\citep{Corbelli1988, Parravano1988, Parravano1989, Ostriker2010, Kim2011, Ostriker2011}.
If the star-formation rate drastically increases,  the high heating rate induced by the stellar UV prevents the formation of the CNM phase, which suppresses  
%AMIEL77
%the star-formation rate.
star-formation.

%AMIEL77 some futrther sliht alterations here. Also linked my added sentence to this paragraph.
Previous computations of the H{\small I} WNM/CNM thermal phase structures have generally assumed that metals dominate the CNM cooling. Then, as would be expected, as the metal abundance is reduced higher particle densities are required to compensate for the reduced cooling, and the multiphase behavior is predicted to occur at higher gas densities and thermal pressures \citep{Wolfire1995, Liszt2002, Walch2011a}.  
The central question we address in this paper is how WNM/CNM multiphase behavior of atomic hydrogen (H{\small I}) gas is affected by H$_2$ cooling and heating processes as the metallicity becomes vanishingly small.  

H$_2$ cooling in the context of primordial and/or low-metallicity gas and first star-formation has been considered by many authors over many years \citep[e.g.,][]{Saslaw1967,
Lepp1984,
Shapiro1987,
Haiman1996a,
Tegmark1997,
Omukai2000,
Abel2000,
Yoshida2006,
Glover2007b,
Schneider2012,
Wolcott-Green2017}.
It has long been known that in the absence of metals, just small amounts of H$_2$ can enable
%AMIEL77 Add the reference to Glover & Clark I have put into this sentence.
cooling to a few 100 K (see for example the recent study by \citealt{Glover2014}) thereby controlling the fragmentation and dynamical properties of collapsing star-forming clouds in the low-metallicity regime. Less attention has been given to the specifics of WNM to CNM transitions and possible multiphase behavior at very low-metallicities. For example, in their study of protogalactic disks \citet{Norman1997} presented pressure versus density thermal phase diagrams for atomic gas over a range of redshifts and associated metallicities. 
In their models
%AMIEL77 added "in their models"
multiphase behavior disappears at high redshifts (see their Fig.~5). However it remains unclear what are the individual roles of the metallicity and radiation field when H$_2$ processes begin playing a role, and how these processes affect the phase structure.  Similarly, \citet{Inoue2015} carried out numerical hydrodynamical simulations of isobarically contracting gas clouds and studied the effects of H$_2$ cooling on thermal instabilities.  They found that at sufficiently low metallicities H$_2$ acts to smooth out thermal instabilities (see their Fig.~2).  But why?

The goal of our paper is to study the competing microphysical processes that control the phase structures of atomic gas, into the regime of vanishing metallicity. We address the following questions. 
How are the WNM and CNM phases  affected by the onset of H$_2$ cooling (and heating)?
Is multiphase gas possible when H$_2$ replaces metal fine-structure line cooling? 
If not, why not? 
In this study we present numerical computations of the (equilibrium) thermal states, with UV radiation intensity, cosmic-ray (or X-ray) ionization rate, and metallicity, as distinct parameters, and with the addition of the H$_2$ processes in a step by step fashion.  
This enables an analytic description of the behavior which we also present.

The structure of our paper is as follows. In \S \ref{sec: model} we present the governing equations, and discuss the chemical and thermal processes we include.
%In \S \ref{sec: Z=1} and \ref{sec: low Z} we present the results.
We begin with a discussion of the solar metallicity model in \S \ref{sec: Z=1}.
We discuss low metallicity behavior in \S \ref{sec: low Z} and \ref{sec: Parameter Study}, down to effectively pristine H$_2$ dominated gas.
We characterize the critical densities or pressures above which the gas may cool below $\sim 10^4$~K and condense into intermediate ($\sim 500$ K) or cold ($\sim 50$~K) phase (depending on metallicity), and whether two-phase media can exist, as function of metallicity, UV intensity and CR ionization rate.
We summarize in \S \ref{sec: conclusions}.

\section{Model Ingredients}
\label{sec: model}

We compute the temperatures and gas phase chemical compositions of the neutral atomic hydrogen (H{\small I}) components of the interstellar medium (ISM) of galaxies
assuming thermal equilibrium and chemical steady state. We use the results to construct pressure versus density phase diagrams for the H{\small I} gas. We consider standard heating and cooling mechanisms, with the addition of molecular hydrogen (H$_2$) processes that become important and eventually dominant in the limit of low metallicity. We investigate the role of the assumed far-UV radiation intensities and cosmic ray-ionization rates on the coupled thermal and chemical properties of the H{\small I} gas. We study the behavior of the thermal phase structures for standard ISM conditions into the limit of vanishing metallicity.

\subsection{Basic Equations}
\label{sub: basic equations}

In thermal equilibrium, 
\begin{equation}
\label{eq: cooling-heating}
L(n,T; x_i) =  G(n,T; x_i)  \ ,
\end{equation}
where $G$ and $L$ are the total heating and cooling rates per unit volume (erg cm$^{-3}$ s$^{-1}$). The rates depend on the total hydrogen volume density $n$ (cm$^{-3}$), the gas temperature $T$ (K), the dust-to-gas abundance ratio, and the gas phase abundances of the chemical species, $x_i \equiv n_i/n$, especially for C$^+$, O, electrons, and H$_2$.
For heating we include far-UV photoelectric (PE) ejection of electrons from dust grains and polycyclic aromatic hydrocarbons  (PAHs),
and cosmic-ray (CR) and/or X-ray ionization of the hydrogen atoms. For cooling, we include H{\small I} Ly$\alpha$ line-emission excited by electron impacts, and fine-structure \cii (157.7$\mu$m), \oi (63.2$\mu$m, 145.5$\mu$m),  and [C{\small I}] (609.1$\mu$m, 370.4$\mu$m) line emissions excited primarily by collisions with the neutral hydrogen atoms. We also include energy losses due to electron recombination with dust grains and PAHs, and due to gas-dust collisions.
%Shmuel5: PAHs also here, above
As discussed by \citet[][hereafter, \citetalias{Wolfire2003}]{Wolfire2003} these are the dominant heating and cooling mechanisms for thermally stable and unstable warm/cold ($\sim10^4$ to $\sim100$ K) interstellar H{\small I} for standard ISM conditions 
%shmuel6: added footnote on metastable cooling
\footnote{Metastable line emission (e.g.~[OI] $6302$ ${\rm \AA}$) also cools the WNM, but its contribution is subdominant (\citetalias{Wolfire2003}, see also Fig 30.1 in \citealt{Draine2011}).}.

Importantly, we also incorporate heating and cooling processes involving H$_2$. 
This includes heating via the release of binding energy during H$_2$ formation, kinetic energy imparted to the H-atoms produced by H$_2$ photodissociation, and internal H$_2$ energy released to the gas via collisional deexcitation of UV excited states (we refer to this latter process as "UV pumping heating"). Cooling is mainly by H$_2$ ro-vibrational line-emissions excited by collisions with electrons and neutrals.
We also include cooling by 
isotopic HD line emissions, and due to dissociation of the H$_2$ in collisions with thermal electrons and H-atoms, 
although these are minor.
In the Appendix we present detailed discussions and expressions for the various heating and cooling processes we have included in our computations.
The roles and effects of H$_2$ heating and cooling processes have not been considered in previous analytic studies of the multiphase structure of the neutral ISM (\citealt{Field1969, Wolfire1995, Liszt2002}; \citetalias{Wolfire2003}).
We will show how the H$_2$ heating and cooling processes influence the thermal structures as the heavy element abundances become small.

Eq.~(\ref{eq: cooling-heating}) is coupled to the equilibrium rate equations for the chemical abundances,
\begin{equation}
\label{eq: chem rates}
\sum_{j>l} x_j x_l \ n k_{jl,i}  +  \sum_{j} x_j q_{j,i} \ = \ 0 \ .
\end{equation} 
This is a set of $N$ equations, equal to the number of species in the chemical network.
%( a_{j,i} \zeta_{\rm cr} +   b_{j,i} I_{\rm UV}) \ = \ 0 \ .
%where the sums,  $j,k=1,2,...,N$, are over all considered species in the chemical network.
The first sum is over the formation and destruction rates (s$^{-1}$) of species $i$ via two-body reactions involving species $j$ and $l$, where
$k_{jl,i}$ are the temperature-dependent
reaction rate coefficients (cm$^3$ s$^{-1}$).
The second sum is over the formation and destruction rates of species $i$ due to ionization or dissociation of species $j$ by either cosmic-rays or UV photons, and the $q_{j,i}$ are the rates (s$^{-1}$) for these processes. 
For $j=i$, the coefficients $k_{jl,i}$ and $q_{j,i}$ are negative and represent the destruction of species $i$ in reactions with species $l$, and by UV and CR dissociations and ionizations. For $j\neq i$ the coefficients are positive and represent the formation channels for species $i$.
Eqs.~(\ref{eq: chem rates}) are augmented by the (dependent) conservation equations for the total elemental abundances and electric charge.
These are
\begin{equation}
\label{eq: conserv}
    X_{m} = \sum_i \alpha_{i, m} x_{i} \ , 
\end{equation}
where $X_{m}$ is the total elemental abundance of element $m$, and $\alpha_{i, m}$ is the number of atoms of element $m$ in species $i$.
Eq.~(\ref{eq: conserv}) also describes the electric charge conservation equation, with $\alpha_{i, m}$ representing the number of positive charges in species $i$, and with $X_{m}=0$.
In total, there are $M+1$ conservation equations, where $M$ is the number of elements in the chemical network.
We discuss our chemical network in \S \ref{sub: numerical method}.

\subsection{Parameters}
\label{sub: params}

The gas temperature $T$, and species abundances $x_i$, depend on the gas density, $n$,  the far-UV radiation intensity, $I_{\rm UV}$, the CR ionization rate, $\zeta_{\rm cr}$, and the metallicity $Z'$. We define these parameters below. Additional physical parameters affecting the thermal and chemical solutions are listed in Table \ref{table: params}.

\begin{table*}[]
\caption{Parameters}
\centering % used for centering table
\begin{tabular}{l l l l}
\hline\hline %inserts double horizontal lines
notation  & fiducial value /  range & meaning & comments
  \\ [0.5ex]
   \hline 
$Z'$ & from 10$^{-5}$ to 3 & metallicity relative to Galactic & see \S \ref{sub: params: Z and DGR}, \S \ref{sec: low Z} \\
$I_{\rm UV}$ &  0.1, 1, 10  & the 6-13.6 eV UV intensity, relative to Galactic & see \S \ref{sub: params: IUV}, \S \ref{sub: var UV zeta} \\
$\zeta_{-16}=\frac{\zeta_{cr}}{10^{-16} {\rm s^{-1}}}$ &  0.1, 1, 10 & the total (primary+secondary) ionization rate per H atom  &  see \S \ref{sub: params: zeta}, \S \ref{sub: var UV zeta} \\
     \\[0.2ex]
   \hline 

   $A_{\rm C}$, $A_{\rm O}$ & 1.4 and 3.2 $\times 10^{-4}$   & the carbon and oxygen solar elemental abundances  & see \S \ref{sub: params: Z and DGR} and Eq.~(\ref{eq: C and O abundances}) \\
      $\delta_{\rm C}$, $\delta_{\rm O}$ & 0.53 and 0.41   & the carbon and oxygen dust depletion factors & see \S \ref{sub: params: Z and DGR} and Eq.~(\ref{eq: C and O abundances}) \\
$Z'_d$ &  & the DGR relative to Galactic & see \S \ref{sub: params: Z and DGR} and Eq.~(\ref{eq: Zd-Z}) \\
$\zeta_p$ &  $\zeta_{cr}/(1+\phi_s)$ & the {\it primary} ionization rate per H atom & see \S \ref{sub: params: zeta} and Eq.~(\ref{eq: phi_s})  \\
$\phi_s$ &  $\approx 0.6$ & the number of secondary electrons produced per ionization & see \S \ref{sub: params: zeta} and Eq.~(\ref{eq: phi_s})  \\

$D_0$ &  $5.8 \times 10^{-11}$~s$^{-1}$ & the H$_2$ photodissociation rate at $I_{\rm UV}=1$  & see \S \ref{sub: params: IUV}, \S \ref{sub: var UV zeta}  \\
$D_-$ &  $5.6 \times 10^{-9}$~s$^{-1}$ & the H$^-$ photodetachment rate at $I_{\rm UV}=1$  & see \S \ref{sub: params: IUV}, \S \ref{sub: var UV zeta}  \\
%$\eta$ &  & the branching ratio for H$^-$ $\rightarrow$ H$_2$ versus photodetachment & see \S \ref{sub: params: R_form}, Eq.~(\ref{eq: eta}) \\
$R$ &  (cm$^{3}$ \ s$^{-1}$) & the H$_2$ formation rate coefficient  & see \S \ref{sub: params: R_form}\\
$x_i=n_i/n$ &  & the species abundance, relative to the hydrogen nucleon density  & see \ref{sub: basic equations} and \ref{sub: params: R_form} \\
$L$, $G$ &  (erg \ cm$^{-3}$ \ s$^{-1}$) & the cooling and heating rates per unit volume  & see \S  \ref{sub: basic equations} and \ref{sub: app_heat_cool}
\\
$\mathcal{L}_i(T)$ &  (erg \ cm$^{3}$ \ s$^{-1}$) & the cooling efficiency of species $i$. For $n/n_{\rm crit} \ll 1$, $L_i=x_i n^2 \mathcal{L}_i(T)$ & see \S \ref{sub: app_heat_cool} \\
$\Lambda_i(T)$ &  (erg \ s$^{-1}$) & the cooling rate per particle. 
For $n/n_{\rm crit} \gg 1$, $L_i=x_i n \Lambda_i(T)$ &  see \S \ref{sub: app_heat_cool} \\
\hline %inserts single line
%\caption*{}
\end{tabular}
\label{table: params} % is used to refer this table in the text
\end{table*}

\subsubsection{Metallicity and dust-to-gas ratio}
\label{sub: params: Z and DGR}
% Shmuel5: changed this subsection

The metallicity, i.e.~the gas phase abundances of the heavy elements, and the dust abundance, are crucial parameters for the thermal balance.
The gas phase carbon and oxygen abundances determine the efficiency of metal-line cooling, and the dust-to-gas mass ratio determines the efficiency of PE heating and dust recombination cooling.
We assume that the total {\it gas-phase} abundances of the carbon and oxygen species are given by
\begin{equation}
\label{eq: C and O abundances}
    X_{\rm m} = A_{\rm m} Z' \left( 1- \delta_{\rm m} \times (Z'_d/Z') \right)  \ ,
\end{equation}
where $m=$C, or O.
% \begin{align}
% \label{eq: C and O abundances}
%     X_{\rm C} &= A_{\rm C} Z' \left( 1- \delta_{\rm C} (Z'_d/Z') \right) \\ \nonumber
%     X_{\rm O} &= A_{\rm O} Z' \left( 1- \delta_{\rm O}(Z'_d/Z') \right) \ \ \ .
% \end{align}
In these expressions $Z'$ is the normalized metallicity such that $Z'=1$ corresponds to the Solar elemental abundances for which $A_{\rm O}=5.4 \times 10^{-4}$ and $A_{\rm C}=3.0 \times 10^{-4}$ \citep{Asplund2009}, $\delta_{\rm C}$ and $\delta_{\rm O}$ are dust depletion factors, and $Z'_d$ is the normalized dust-to-gas ratio.
We adopt
\begin{equation}
\label{eq: Zd-Z}
Z_d' = 
\begin{cases}  Z' & \text{for } Z' \geq Z_0' \\
 Z_0' \ (Z'/Z_0')^{\alpha} & \text{for } Z' < Z_0' \ ,
\end{cases}
\end{equation} 
with $Z_0'=0.2$, and $\alpha=3$. 
This broken power-law relation is suggested by recent observations of low metallicity galaxies \citep{Remy-Ruyer2013}. Here, $Z'_d=1$ corresponds to a standard ISM dust-to-gas ratio of $6.2 \times 10^{-3}$ \citep[][Table 6, the BARE-GR-S model]{Zubko2003}. 

% For this relation the dust grain heating and cooling efficiencies vanish rapidly as the metallicity is reduced. 
For the dust depletion factors, we adopt
$\delta_{\rm C}= 0.53$ and $\delta_{\rm O}= 0.41$. 
With these values,  we recover the characteristic interstellar Galactic carbon and oxygen gas-phase abundances $X_{\rm C} = 1.4 \times 10^{-4}$, $X_{\rm O}=3.2 \times 10^{-4}$ at $Z'=Z'_d=1$ \citep{Cardelli1996, Sofia1997, Meyer1997}.
On the other hand, at very low metallicities, when $Z'_d/Z' \ll 0.2$, the dust depletion terms vanish, and all the available carbon and oxygen remain in the gas-phase, with $X_{\rm C} = A_{\rm C}Z'$, $X_{\rm O}=A_{\rm O}Z'$.

% We scale the elemental abundances with a common factor $Z'$, such that $Z'=1$ corresponds to solar metallicity. For $Z'=1$ the dust-to-gas mass ratio equals XXX (REF).
% In particular, we assume that
% \begin{equation}
% \label{eq: Zd-Z}
% Z_d' = 
% \begin{cases}  Z' & \text{for } Z' \geq Z_0 \\
%  Z_0 \ (Z'/Z_0)^{\alpha} & \text{for } Z' < Z_0 \ ,
% \end{cases}
% \end{equation} 
% with $Z_0'=0.2$, and $\alpha=3$. 
% This broken power-law relation is suggested by recent observations of low metallicity galaxies \citep{Remy-Ruyer2013}. 
% For this relation the dust grain heating and cooling efficiencies vanish rapidly as the metallicity is reduced. 
%SHMUEL: below $Z~0.1$ 

\subsubsection{Cosmic-ray and X-ray ionization}
\label{sub: params: zeta}
Cosmic rays (or X-rays) ionize atomic and molecular hydrogen, 
\begin{align}
\label{reac: H CR}
{\rm H \ + \ cr \ } & \rightarrow {\rm \ H^+ \ + \ e} \\ 
{\rm H_2 \ + \ cr \ } & \rightarrow {\rm \ H_2^+ \ + \ e} \\ 
{\rm H_2 \ + \ cr \ } & \rightarrow {\rm \ H^+ \ + \ H \ + \ e}
\end{align}
producing energetic non-thermal electrons.
The electrons generated by these primary ionizations lose energy through coulomb scatterings that heat the gas, and through further atomic and molecular excitations and secondary ionizations \citep{Dalgarno1999a}. 
The total (primary+secondary) ionization rate per hydrogen atom is $\zeta_{\rm cr}=\zeta_p \times (1+\phi_s)$, where $\zeta_p$ is the primary ionization rate and $\phi_s \sim 0.7$ is the number of secondary electrons produced per primary ionization.
The cosmic-ray (CR) ionization rate determines the CR heating rate
and the electron abundance (see Eqs.~\ref{eq: xe} and \ref{eq: CR heating}, below).
For typical Galactic values, $\zeta_{\rm cr} \approx 10^{-16}$~s$^{-1}$ \citep{Indriolo2012, Tielens2013}, and we define
the normalized CR ionization rate, $\zeta_{-16} \equiv \zeta_{\rm cr}/(10^{-16} \ {\rm s^{-1}})$.
Our fiducial value is $\zeta_{-16} =1$. We study the effects of varying $\zeta_{-16}$ in \S \ref{sub: var UV zeta}. 

% The cosmic-ray (CR) ionization rate determines the gas heating rate and the electron abundance, $x_{\rm e}$.
% We discuss the 
% Cosmic-ray heating is proportional to the primary ionization rate $\zeta_p$ (see Eq.[A4]). In general, $\zeta_p = \zeta_{\rm cr}/(1+\phi_s)$, where $\phi_s$ is the number of secondary electrons produced per primary ionization. Following \citet[][Eq.~(13.12)]{Draine2011}, we assume
% \begin{equation}
% \label{eq: phi_s}
% \phi_s = \left( 1 - \frac{x_{\rm e}}{1.2} \right) \frac{0.67}{1+(x_{\rm e}/0.05)} \ \ \ .
% \end{equation}

X-rays can play an important role in providing an additional source of  ionization, affecting the chemistry and the heating rate.
We do not explicitly include X-ray ionization in this study, 
however, 
% the forms for the X-ray ionization and heating rates are similar to the cosmic-ray processes, 
these may be simulated by choosing an appropriate values for $\zeta_{\rm cr}$ and $\phi_s$.

\subsubsection{Interstellar far-UV radiation field}
\label{sub: params: IUV}

The far-UV radiation intensity determines the efficiency of photoelectric heating, the H$_2$ photodissociation rate, and the production rate of the C$^+$ coolant.
\citet{Draine1978, Draine2011} estimated the local UV interstellar radiation field (ISRF), and obtained 
$u_{\rm UV,0}=0.056$ eV cm$^{-3}$, for the energy density over the  6-13.6 eV band (=1.69 times the \citealt{Habing1968} estimate).
We adopt the Draine spectral shape and allow for variations in the UV intensity by multiplying the overall spectrum with a normalized intensity parameter, $I_{\rm UV}$, such that $u_{\rm UV} = u_{\rm UV,0} I_{\rm UV}$.
The UV intensity determines the PE heating rate through Eq.~(\ref{eq: p.e. heating}).
For a unit Draine ISRF, the free-space (unattenuated) H$_2$ photodissociation rate is
$D_0= 5.8 \times 10^{-11}$~s$^{-1}$ \citep{Sternberg2014}. 
%We discuss UV attenuation below.
At low metallicity, a harder spectrum may be more typical, since the first generations of stars (pop-III stars) are theorized to be very massive, with stellar effective temperatures as high as 10$^5$ K \citep[e.g.,][]{Cojazzi2000, Barkana2001, Abel2002, Yoshida2012}.
% The advantage of choosing the Draine spectrum for all metallicities is that the UV density, $u_{\rm UV}$, and the H$_2$ photodissociation rate, $D_0$, converge to the observed Galactic values at solar metallicity.
% In \S \ref{sub: var UV zeta}, we present models for a range of assumed values for $I_{\rm UV}$.
%and discuss the effects of variations to the H$_2$ photodissociation rate. 
%At low $Z'$ such variations are also equivalent to varying the spectrum shape, since then UV only plays a role in H$_2$ photodissociaiton, not in PE heating.

Longer wavelength photons, down to 0.75 eV, play a role in removing the H$^-$ anion 
via photodetachment \citep{Miyake2010}. H$^-$ is a critical intermediary for H$_2$ formation at low metallicity (see \S \ref{sub: params: R_form}, below).
\citet{Mathis1983} estimated the Galactic ISRF within this band as a combination of diluted blackbodies at 3000, 4000 and 7500 K.
For an ISRF combined of the Draine field over the 6-13.6 eV band, and with the Mathis field over the 0.75-6 eV band, the H$^-$ photodetachment rate is $D_- =2.7 \times 10^{-7}$ s$^{-1}$.
In contrast, for a 10$^5$ K blackbody (BB) spectrum (normalized such that the photon density in the 6-13.6 eV band equals that of the Draine field), $D_- = 2.8 \times 10^{-9}$ s$^{-1}$ (\citealt{Bialy2015a}; see their Table 2).
For our fiducial models we seek a compromise between these two extreme cases and extrapolate the \citet{Draine1978} spectrum (see their Eq.~11) down to 0.75 eV. 
The resulting H$^-$ photodetachment rate is $D_- = 5.6 \times 10^{-9}$~s$^{-1}$.
This value is close to the 10$^5$ K BB value which represents the pristine metallicity gas, but is $\sim 2$ orders of magnitude smaller than the Draine+Mathis value, which represents the Galaxy and a high metallicity ISM.
For moderately low metallicities $D_-$ may range between these two extremes, and will depend on the form of the initial mass function (IMF).
We explore the effect of variations in $I_{\rm UV}$ and $D_-$ in \S \ref{sub: var UV zeta}.

%shmuel6: removed the subsection on Radiation Attenuation. 
% Added this text instead
The free-field photodissociation rate, $D_0$ may be strongly reduced in the interiors of a cloud due to 
dust absorption and H$_2$ self-shielding, leading to partial or complete H {\small I}-to-H$_2$ conversion \citep[][]{VanDishoeck1986, Draine1996, McKee2010, Sternberg2014, Bialy2016a}.
In this paper we focus on the thermal properties of the {\it global atomic} ISM, rather than the structure of the H {\small I}-to-H$_2$ transition or the interior shielded molecular gas.
Thus, we always use the free-space UV field and do not include radiation attenuation in our models.
For the WNM shielding is in any case not significant (except for extremely large shielding columns).
For the CNM, our model applies to the cloud outskirts that are exposed to the free-field UV.

\subsubsection{H$_2$ Formation}
\label{sub: params: R_form}

At sufficiently high metallicity, the H$_2$ is formed predominantly on dust grains.
We adopt 
\begin{equation}
\label{eq: RZ}
R_{d} =3 \times 10^{-17} \left( \frac{T}{ \rm 100 \ K }\right)^{1/2} Z_d' \ \ {\rm cm}^3 {\rm s}^{-1} \ \ \ .
\end{equation}
Sophisticated models of the H$_2$ formation process on  dust-grains \citep[e.g.,][]{Cazaux2002,LeBourlot2012, Bron2014} derived more complicated dependences on gas temperature, as well as on other physical parameters (e.g., through the dust temperature which in turn depends on the UV radiation intensity). However for our purposes such a simple description is sufficient, because as we show below, H$_2$ affects the thermal structure only at low metallicities where its formation is dominated through gas-phase processes.

% \begin{equation}
% \label{eq: RZ}
% R_{\rm dust} = 3\times 10^{-17} T_2^{1/2} \left( \frac{S_H(T)}{1+S_H(T)} + \mathrm{e}^{-T_2/3} \right)Z_d'  \ \ {\rm cm^3 \ s^{-1}} ,
% \end{equation}
% for the H$_2$ formation rate coefficient on dust grains
% where $T_2 \equiv T/(10^2 \ {\rm K})$, and $S_H = \mathrm{e}^{-T_B/T}/(1+[T/T_0]^{\beta})$
%\begin{equation}
%S_H \equiv \frac{\mathrm{e}^{-T_B/T}}{1+(T/T_0)^{\beta}} \ 
%\end{equation}
% is the sticking coefficient. We assume $T_B=300$~K, $T_0=464$~K, $\beta=1.5$ \citep{LeBourlot2012}. We have modified the original \citet{LeBourlot2012} expression by adding an exponential term %
% %$\mathrm{-T/300 {\rm K}}$%
% to $S_H/(1+S_H)$ such that $R$ does not vanish at $T \lesssim 100$~K, but remains $\approx 10^{-17}$~cm$^3$~s$^{-3}$.

 At low metallicities ($Z' \lesssim 0.1$), H$_2$ is formed mainly through gas-phase reactions, with the most important one being the H$^-$ formation route \citep{McDowell1961, Peebles1968, Hirasawa1969, Galli1998a}.
 Although, the H$_2$ formation rate is obtained through the numerical solution to the chemical network (as discussed in \S \ref{sub: basic equations}), it is useful to obtain approximate analytic equations for the effective gas-phase formation rate coefficient, and the resulting H$_2$ fractional abundance.
 
This formation channel is initiated by radiative attachment
\begin{equation}
\label{reac: H- formation}
{\rm H \ + \ e \ \rightarrow \ H^- + \nu} \ ,
\end{equation}
followed by associative detachment
\begin{equation}
\label{eq: H2 form gas reac}
{\rm H^{-} \ + \ H \ \rightarrow \ H_2 \ + \ e } \ \ \ .
\end{equation}
Assuming that the electron abundance is set by a balance between CR ionization of the atomic hydrogen and proton-electron recombination, we have
\begin{equation}
\label{eq: xe}
x_{\rm e} = \left( \frac{\zeta}{\alpha_B n} \right)^{1/2} = 2.7 \times 10^{-3} T_3^{3/8}  \zeta_{-16}^{1/2} \left( \frac{n}{10 \ {\rm cm^{-3}}} \right)^{-1/2}\ ,
\end{equation}
where $T_3 \equiv T/({\rm 10^3 \ K})$
% $\alpha_B=1.4\times 10^{-12}T_3^{-0.75}$~cm$^{3}$~s$^{-1}$
and $\alpha_B$ is the case-B recombination rate coefficient.
The effective H$_2$
formation rate coefficient through this channel is
\begin{equation}
\label{eq: R gas phase}
R_{\rm -} \equiv x_{\rm e} k_{\ref{reac: H- formation}} \eta \approx 1.9 \times 10^{-18} \  T_3^{1.02} \eta \ \zeta_{-16}^{1/2} \left( \frac{n}{10 \ {\rm cm^{-3}}} \right)^{-1/2} \ {\rm cm^3 \ s^{-1}}
\end{equation}
%\begin{align}
%\label{eq: R gas phase}
%R_{\rm -} &= x_{\rm e} k_- \eta = \sqrt{\frac{\zeta}{\alpha_B n}} k_- \eta \\ \nonumber
%& \approx 5.4 \times 10^{-19} \  T_2^{1.02} \eta \ \zeta_{-16}^{1/2} n^{-1/2} \ {\rm cm^3 \ s^{-1}}
%\end{align}
 where 
$k_{\ref{reac: H- formation}}$
% =7.2 \times 10^{-16} T_3^{0.64} \ {\rm e}^{-9.2 {\rm K}/T}$ ~cm$^{3}$~s$^{-1}$
is the radiative attachment rate coefficient\footnote{The rate coefficients are based on UMIST 2012 \citep{McElroy2013}.
See Table \ref{table: reduced netwrok} for a summary of the most important reactions and rates. See online material for the full network.} (Eq.~\ref{reac: H- formation}).
The H$_2$ formation sequence is moderated mainly by photodetachment,
\begin{equation}
\label{reac: H- photodetachment}
{\rm H^- \ + \nu \ \rightarrow \ H \ + \ e} \ \ \ .
\end{equation}
and also by mutual neutralization with protons,
\begin{equation}
\label{reac: mutual neutralization}
{\rm H^- \ + H^+ \ \rightarrow \ H \ + \ H} \ \ \ .
\end{equation}
In Eq.~(\ref{eq: R gas phase}) we define the branching ratio
\begin{equation}
\label{eq: eta}
\eta = \left( 1+ \frac{D_- I_{\rm UV}}{k_{\ref{eq: H2 form gas reac}} n} + \frac{k_{\ref{reac: mutual neutralization}} x_{\rm H^+}}{k_{\ref{eq: H2 form gas reac}}} \right)^{-1} 
\end{equation}
for H$_2$ formation via associative detachment versus H$^-$  photodetachment and mutual-neutralizaion.
% We now obtain an analytic approximation for the H$_2$ fractional abundance at low metallicities.
For $D_-=5.6\times 10^{-9}$ s$^{-1}$ 
% and with $k_{\ref{reac: mutual neutralization}}=4.1\times 10^{-8} T_3^{-0.5}$ cm$^3$ s$^{-1}$, $k_{\ref{eq: H2 form gas reac}} = 2.6 \times 10^{-9} T_3^{-0.39} {\rm e}^{-39.4 {\rm K}/T}$ cm$^3$ s$^{-1}$, we obtain that
photodetachmnet becomes important when
 $n/I_{\rm UV} \lesssim 1$~cm$^{-3}$, while mutual-neutralization 
 is important when
$n/\zeta_{-16} \lesssim 0.01$~cm$^{-3}$.
As we show below, H$_2$ plays an important role in cooling and heating at densities $\gg 1$ cm$^{-3}$.
Thus, generally, $\eta \approx 1$ is a good approximation \footnote{For high $D_-$ values, $\eta$ may become smaller than unity at the cooling point, reducing the H$_2$ abundance. We discuss this in \S \ref{sub: var UV zeta}.}. 

Assuming that H$_2$ destruction is dominated by photodissociation, 
% and using Eq.~(\ref{eq: R gas phase}) for $R_-$
% with $\eta=1$, 
we obtain an analytic approximation for the H$_2$ fractional abundance at low metallicities,
\begin{equation}
\label{eq: xH2 gas}
x_{\rm H_2} =  \frac{x_{\rm e} k_{\ref{reac: H- formation}} n}{D_0 I_{\rm UV}}  \eta = 3.3 \times 10^{-7} T_3^{1.01} \eta \zeta_{-16}^{1/2}I_{\rm UV}^{-1} \left( \frac{n}{10 \ {\rm cm^{-3}}} \right)^{1/2}  \ .
\end{equation}
% where we assume H$_2$ removal through photodissociation by the free-space UV radiation. 
% For $I_{\rm UV}=1$ and $T=100-1000$~K, this expression gives $x_{{\rm H_2}}$ ranging from XXX to XXX for gas densities $n$ between 10$^2$ and 10$^4$ cm$^{-3}$, and consistent with the numerical results we present below.
We find Eq.~(\ref{eq: xH2 gas}) to be in good agreement with the numerical solution for the chemical network, over the gas densities and temperatures in our parameter space.

\begin{figure*}[t]
	\centering
	\includegraphics[width=1\textwidth]{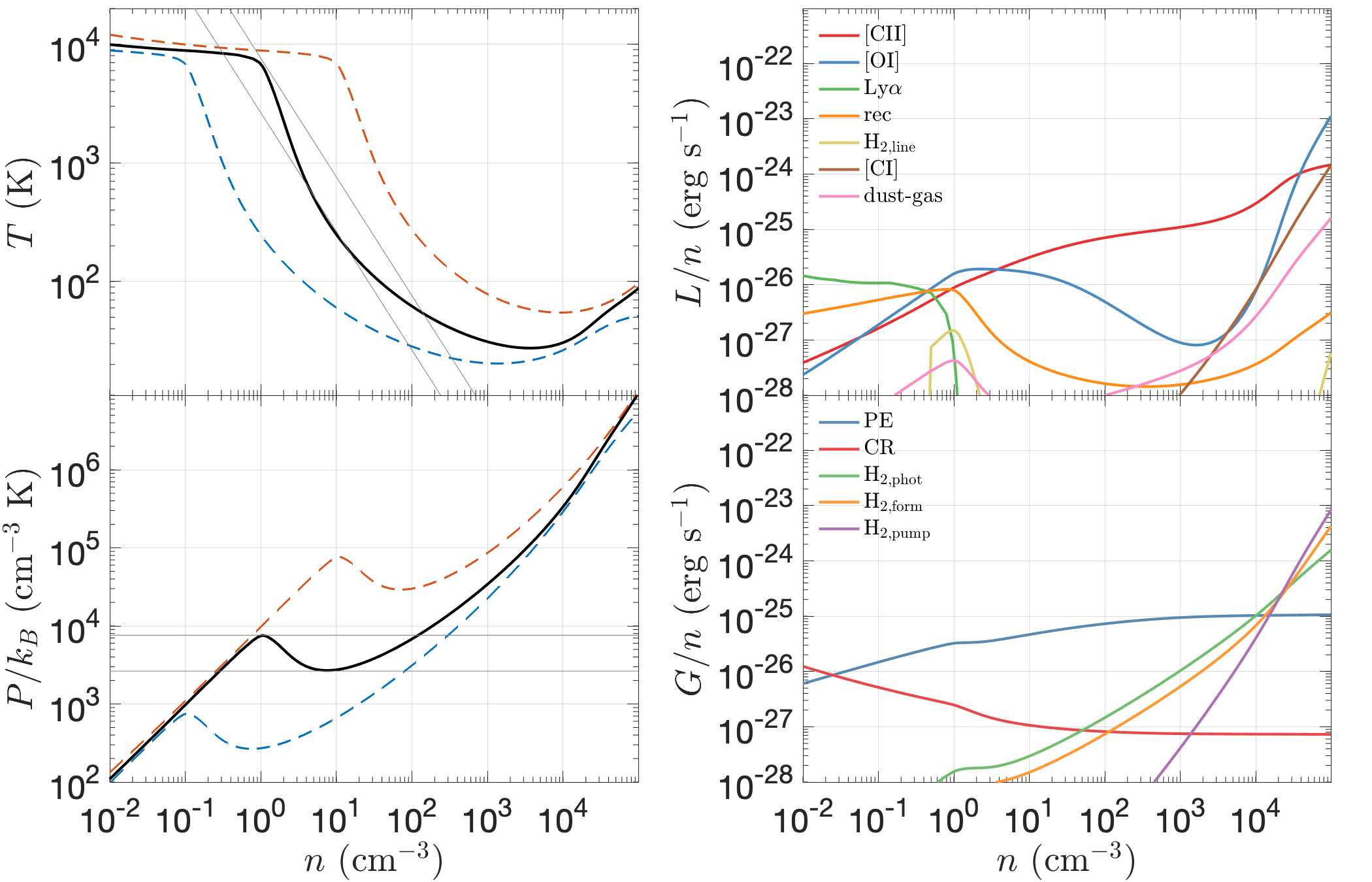} 
	\caption{
The temperature, pressure, and cooling and heating rates per hydrogen nucleus as functions of the volume density, assuming solar metallicity and $I_{\rm UV}=\zeta_{-16}=1$. 
The thin diagonal (horizontal) lines in the left panels indicate the minimum and maximum pressures for a multiphase medium.
The dashed red and blue curves in the left panels show models with $I_{\rm UV} = \zeta_{-16} = 0.1$ and 10, respectively.
		}    %/Turb//plots_for_paper/profiles_and_N1_tot_vs_
		\label{fig: Tn_Z1}
\end{figure*}

\subsection{Numerical Method}
\label{sub: numerical method}

Given a set of parameters, $I_{\rm UV}$, $\zeta_{\rm cr}$, and $Z'$, we solve for the equilibrium temperature and species abundances as functions of the hydrogen density, $n$.
We form a logarithmic grid, spanning $\log T=1$ to $4.2$ and $\log n=-2$ to 6, with $100 \times  90$ sampling points.
For each volume density in the grid, we calculate the species abundances coupled with the net cooling, $L-G$, as functions of the temperature, and find the temperature(s) for which the thermal equilibrium condition (Eq.~\ref{eq: cooling-heating}) is satisfied.
This gives us the $T(n)$ curves shown in Figs.~\ref{fig: Tn_Z1} and \ref{fig: Tn_diff_Z} below. We compute the equilibrium thermal pressure (cm$^{-3}$~K) by simply multiplying by the density, i.e., $P(n)=nT(n)$. For any density, heating exceeds cooling for pressures less than $P(n)$, and cooling exceeds heating for pressures greater than $P(n)$.

%The net cooling is the sum of all heating and cooling rates, as discussed in \S %\ref{sub: basic equations} and in the Appendix.
For our chemical network (Eq.~\ref{eq: chem rates}),  
we consider a set of 294 two-body gas-phase reactions among 34 species, composed of hydrogen, helium, carbon, and oxygen bearing atomic and molecular species.
We adopt the rate coefficients suggested by the UMIST 2012 database \citep{McElroy2013}, except for the H$_2$ collisional dissociation for which we adopt the \citet{Martin1996} rate (including both collisional dissociation and dissociative tunneling). 
We also include the  formation of H$_2$ on dust grains, CR reactions and photo-reactions.
The full reaction list is provided in the supplement materials (online version).
The dominant photorates are also listed in \citealt[][see their table 2]{Bialy2015a}.

In total we obtain a set of 34 coupled non-linear algebraic equations for the 34 species. The equations depend on $T$ and $n$, as well as on the additional physical parameters, $I_{\rm UV}$, $\zeta_{\rm cr}$, and $Z'$.
%\footnote{The equilibrium abundances are in most cases unique. 
%For some regions in the parameter space
%although in some cases there may exist more than a one set of possible solutions for abundances. However, this phenomena occurs within a typically narrow range of the parameter space that we consider, see (cite XXX) for a discussion of these chemical instabilities. }.
We replace five of these equations with the element conservation (H, He, C, and O), and charge conservation equations. For any density $n$ there is always a single unique solution for the temperature $T$ and abundances $x_i$.  
Rather than solving the equilibrium algebraic equations (e.g., via Newton-Raphson iteration), we instead integrate the time-dependent rate equations until a time, $t$, where $dx_i/dt=0$ and equilibrium is reached.
We find this approach to be more robust. 

To improve computational speed, we also consider a reduced chemical network,
which allows us to calculate the steady-state temperature as a function of metallicity and volume densities (or pressures) over a very well sampled grid, as presented in Figs.~\ref{fig: T_2d} and  \ref{fig: T_2d_vars}.
The reduced network is discussed in the Appendix.

%, where we also present a comparison to the full chemical network, showing excellent agreement over a wide range of metallicities, densities and temperatures.

%For our cause, of calculating the temperature in the predominantly atomic regime, this minimal network provides sufficient accuracy.

%Other dust assited chemical reactions do not play an important role in the atomic ISM, especially not at low metallicities.
%Three-body gas-phase reactions, become important only at large volume densities, beyond the parameter space considered.

\section{Thermal Structure at Solar Metallicity}
\label{sec: Z=1}

In this section we first present results for our solar metallicity reference models ($Z'=Z'_d=1$).
%We assume $A_{\rm He}=0.1$, $A_{\rm C}=3.2 \times 10^{-4}$, $A_{\rm O} = 1.4 \times 10^{-4}$, for the helium, carbon and oxygen elemental abundances, relative to Hydrogen \citep{Wolfire2003}.
%We discuss how the thermal structure depends on the most important physical parameters: the UV intensity, $I_{\rm UV}$, and the CR ionization rate, $\zeta_{\rm cr}$.
%In the following section, \S XXX, we present results for low metallicity gas.
%In both sections we give a special emphasis on the role of H$_2$ heating and cooling.
%We start with a discussion of the fiducial model: $I_{\rm UV}=\zeta_{-16}=1$.
%In \S XXX we discuss the effect of $I_{\rm UV}$ and $\zeta_{-16}$. 
The solid curves in Fig.~\ref{fig: Tn_Z1} show the temperature (upper left panel), the thermal pressure (lower left panel), and the cooling and heating rates per hydrogen nucleus (right panels), as functions of the hydrogen nucleus gas density $n$, assuming $I_{\rm UV}=\zeta_{-16}=1$.
The dashed curves in the left panels, located above and below the solid curve, are for $I_{\rm UV}=\zeta_{-16}=10$ and 0.1, respectively.

%shmuel6: change here
First we focus on the behavior at low to intermediate densities, $n \lesssim 10^3$ cm$^{-3}$, where H$_2$ cooling and heating processes are subdominant.
In this regime, our results
% we show in Fig.~\ref{fig: Tn_Z1} for $T(n)$ and $P(n)$ 
essentially reproduce those presented by \citet{Wolfire1995} and \citetalias{Wolfire2003}. 
The behavior is controlled by the temperature dependence of the Ly$\alpha$ and fine-structure metal-line cooling functions. We display these cooling functions in the upper panel of Fig.~\ref{fig: cooling functions}. The black curve shows the exponentially steep electron impact excitation Ly$\alpha$ cooling rate (erg s$^{-1}$) per hydrogen nucleus. The blue curves show the much flatter metal cooling rates for $Z'=1$, 10$^{-2}$, and 10$^{-4}$, assuming all of the gas-phase carbon is C$^+$, and the oxygen is in neutral atomic form. The metal cooling is controlled by electron and neutral-hydrogen impact excitation of the fine-structure levels. The solid and dashed curves in Fig.~\ref{fig: cooling functions} are for $x_{\rm e}=10^{-4}$ and 10$^{-2}$ respectively. For the high $x_{\rm e}$, and when $T \lesssim 500$ K, metal line cooling is dominated by electron impact excitation. 
% For example, in the fiducial model presented in Fig.~\ref{fig: cooling functions}, at $n=10$ cm$^{-3}$, $T \arpprox 200$ K, where $x_{\rm e} \approx 10^{-2}$
% At densities \GG 10, the electron fraction falls well below 10^{-2}
The shapes of these curves are of the utmost importance.

Cooling becomes inefficient compared to heating as the density becomes small.
For $n \lesssim 0.1$~cm$^{-3}$, the gas becomes warm, and Ly$\alpha$ dominates the cooling.
In this regime the temperature depends weakly on density and remains close to 10$^4$~K.
This is because for $T \lesssim 10^4$~K Ly$\alpha$ cooling is exponentially sensitive to the temperature (see Fig.~\ref{fig: cooling functions}), and any increase in the cooling rate due to an increase in gas density requires just a small decrease in temperature to balance the heating rate. At $n \approx 1$ cm$^{-3}$,
the temperature decreases to $\sim 6 \times 10^3$~K, at which point energy losses by the \cii and \oi emissions become comparable to Ly$\alpha$, as seen in Fig.~\ref{fig: Tn_Z1} and Fig.~\ref{fig: cooling functions}.
Above this density the metals dominate the cooling and the temperature falls abruptly.
The sharp drop occurs because at these high temperatures (and as long as $T \gtrsim 300$~K, i.e.~above the transition energies of the fine-structure lines) metal cooling is a weak function of temperature (Fig.~\ref{fig: cooling functions}). Any increase in density then requires a large decrease in temperature to balance the heating rate.
We refer to the density at which metal cooling takes over as the ``cooling point density", and we provide analytic estimates in \S 4.
Below a few 100~K, the metal line cooling becomes exponentially temperature-sensitive (Fig.~\ref{fig: cooling functions}), and the equilibrium $T(n)$ curve flattens again.

In Fig.~\ref{fig: Tn_Z1}, the low-density flat part of the $T(n)$ curve followed by the sharp temperature drop gives rise to a local maximum, $P_{\rm max}$, in the $P(n)$ curve.
In general, a local maximum such as this occurs if and where the temperature starts decreasing more rapidly than $1/n$.
% as the density is increased.
For our fiducial model with $I_{\rm UV}=\zeta_{-16}=1$, we find $P_{\rm max}=7.6 \times 10^3$ cm$^{-3}$ K occurring at a density $n_{\rm max}=1.1$ cm$^{-3}$.  For $n<n_{\rm max}$ the gas is warm with $T_{\rm WNM}=6000$ to $10^4$~K. This is the classical warm neutral medium (WNM), and $P_{\rm max}$ is the maximum pressure possible for the WNM.
Starting from cold gas at high densities, the steep rise of the $T(n)$ curve as the density is {\it reduced
} gives rise to a local minimum, $P_{\rm min}$, in the pressure versus density curve. 
In general, a local minimum occurs if and where the temperature starts rising more rapidly than $1/n$ as the density is reduced.
For our fiducial model $P_{\rm min}=2.6 \times 10^3$ cm$^{-3}$ K at $n_{\rm min}=7.3$ cm$^{-3}$. For $n>n_{\rm min}$, the gas remains cold with $T_{\rm CNM} \lesssim 300$ K.  This is the cold neutral medium (CNM) and $P_{\rm min}$ is the minimum pressure possible for the CNM.  
We show $P_{\rm min}$ and $P_{\rm max}$ in lower (upper) panels of Fig.~\ref{fig: Tn_Z1} as the pair of thin horizontal (diagonal) lines.

Our Fig.~\ref{fig: Tn_Z1} reproduces the well-known result \citep{Field1969, Wolfire1995} that in the transition from Ly$\alpha$ to metal-line cooling the temperature falls more rapidly than $1/n$ over a significant density interval from $n_{\rm max}$ to $n_{\rm min}$. This enables multiple solutions for the gas temperature between $P_{\rm min}$ and $P_{\rm max}$ for densities ranging from $n_{\rm min}T_{\rm CNM}/T_{\rm WNM}$ to $n_{\rm max}T_{\rm WNM}/T_{\rm CNM}$.  For densities between $n_{\rm max}$ and $n_{\rm min}$ the gas is thermally unstable and this is the UNM. For isobaric density perturbations $\delta n$ in the UNM, the gas either cools to the stable CNM branch (for positive perturbations), or heats to the stable WNM branch (for negative perturbations).  For thermal pressures between $P_{\rm max}$ and $P_{\rm min}$ the gas may become multiphased with CNM condensations coexisting within an enveloping WNM at the same thermal pressure.

%shmuel6: new
Fig.~\ref{fig: Tn_Z1} shows that for $n \gtrsim 5 \times 10^3$ cm$^{-3}$, the temperature 
{\it increases} with density, as H$_2$ heating processes start contributing. 
In this regime the H$_2$ heating rate increases more steeply than $n^2$ (see Eq.~\ref{eq: H2 heating} in the limit $n/n_{\rm crit}<1$) leading to the temperature increase with $n$. 
We stress that the H$_2$ heating processes play this role so long as the gas is primarily atomic. If any self-shielding induces an HI-to-H$_2$ transition, the H$_2$ heating will be strongly suppressed. For solar metallicity a total (atomic plus molecular) column as small as $N\sim 10^{17}$ cm$^{-2}$ is sufficient to induce conversion to molecular form for $n \gtrsim 10^4$ cm$^{-3}$ (and $I_{\rm UV}=1$). However, at low metallicities, much large columns, $N\sim 10^{21}$ cm$^{-2}$, are required \citep{Bialy2016a}

% As long as the density does not exceed the H$_2$ critical density of $\sim 10^5$ cm$^{-3}$ for vibrational deexcitation (following UV pumping) 
At still higher densities (beyond the considered parameter space) the hydrogen undergoes a conversion into H$_2$ and the atomic carbon and oxygen convert into CO and H$_2$O. These molecular formation processes become efficient as the temperature exceeds few hundred Kelvins, which triggers the neutral-neutral chemical network of H$_2$O \citep{Sternberg1995, VanDishoeck2013a, Bialy2015b}.
As our focus is on the (predominantly) atomic ISM, we restrict ourselves to densities $n \lesssim 10^5$ cm$^{-3}$.

The dashed curves in the left panel of Fig.~\ref{fig: Tn_Z1} are the equilibria for $I_{\rm UV}=0.1$ and 10, assuming a constant ratio $\zeta_{-16}/I_{\rm UV}=1$.
For all values of $I_{\rm UV}$ in Fig.~\ref{fig: Tn_Z1} photoelectric heating dominates and is proportional to $I_{\rm UV}$.
% check if this is correct and why.
Thus, when $I_{\rm UV}$ is increased (decreased) by a factor of 10, the $T(n)$ curves shift to the right (left), by a factor of 10.
As a result, the pressure range within which the gas may be multiphased shifts to higher or lower values in proportion to $I_{\rm UV}$. 
For example, for $I_{\rm UV}=10$, $(P_{\rm max}, P_{\rm min})= (7.7, 2.9) \times 10^4$ cm$^{-3}$ K.
For $I_{\rm UV}=0.1$, $(P_{\rm max}, P_{\rm min})= (7.5, 2.7) \times 10^2$ cm$^{-3}$ K.

% At the high density end, the temperature reaches $T = 1300$ and 120 K, for $I_{\rm UV}=10$ and 0.1, respectively.
% For the high UV case, this is in good agreement with Eq.~(\ref{eq: T_plat}) which predicts 1400 K.
% For the low UV case, the temperature is lower than given by Eq.~(\ref{eq: T_plat}) because at this weak UV intensity the heating rate is low enough and is balanced by metal cooling, not by H$_2$ cooling (as assumed in Eq.~\ref{eq: T_plat}).

At solar metallicity, H$_2$ cooling and heating become important only in the very dense CNM, 
but as we discuss below, at lower metallicities, H$_2$ cooling starts to play a role also at low-to-intermediate densities, in the WNM, affecting the WNM-to-CNM transition.

\section{Thermal Structure at Low Metallicity}
\label{sec: low Z}

In this section we investigate the thermal structures as the metallicity is lowered. We first consider the transition from photoelectric to cosmic-ray dominated heating as $Z'$ is reduced, for a range of ratios $\zeta_{-16}/I_{\rm UV}$. We then solve the equation of thermal equilibrium, in two steps. First, we lower the metallicity and solve for $T$ with the exclusion of the H$_2$ heating and cooling processes. Then we solve again with the H$_2$ included. This procedure enables us to distinguish the effects of diminished metal line cooling from the appearance of the H$_2$ heating and cooling processes that eventually dominate the behavior at very low metallicity.

\begin{figure}[t]
	\centering
	\includegraphics[width=0.5\textwidth]{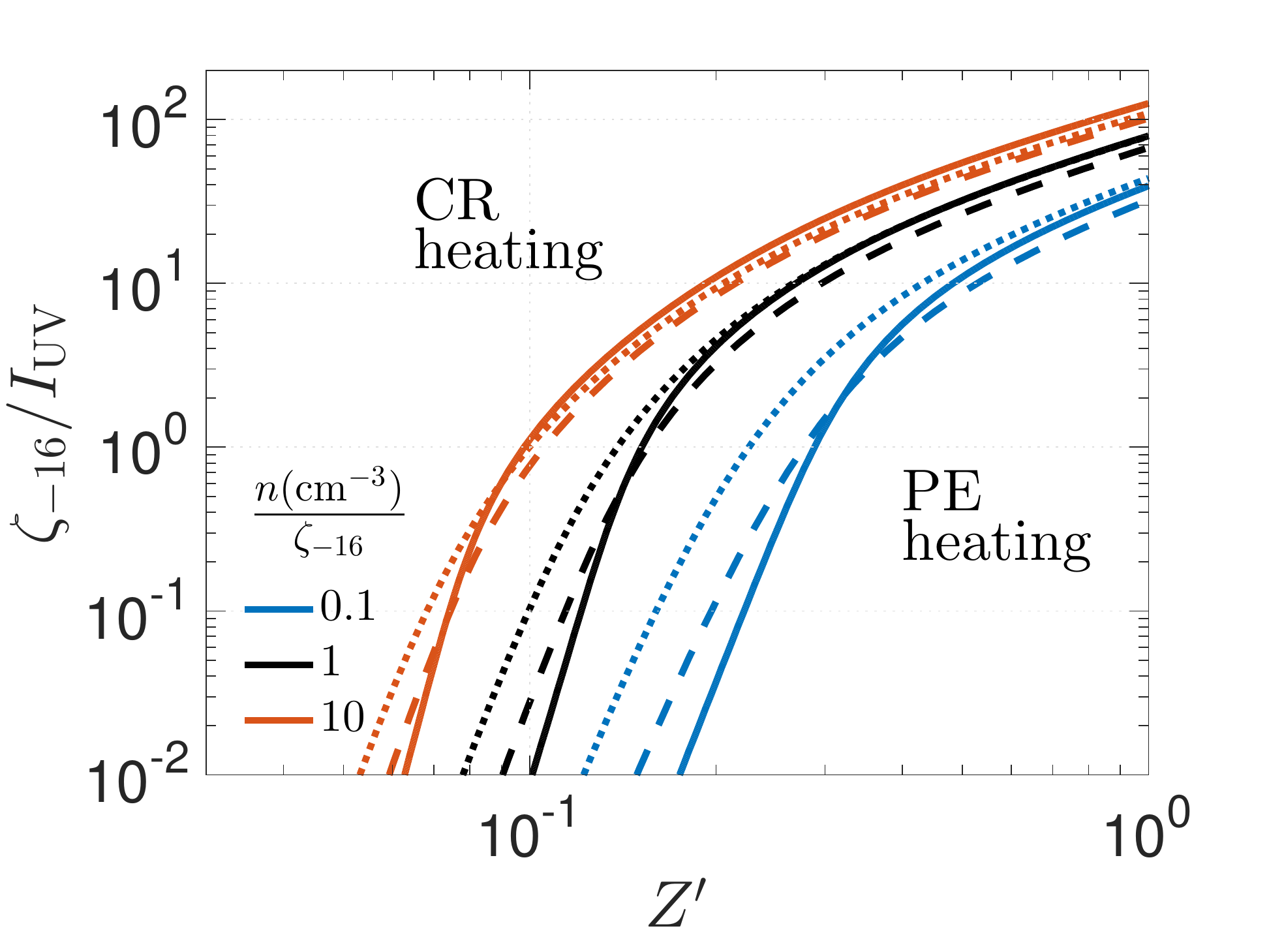} 
	\caption{
Contours at which CR heating equals PE heating in the $I_{\rm UV}/\zeta_{-16}-Z'$ parameter space. The solid, dashed and dotted curves are for $T=10^4, 10^3$ and 10$^2$ K respectively, and the different colors correspond to different values of $n/\zeta_{-16}$, from  $n/\zeta_{-16}=10$ (left most curves) to $n/\zeta_{-16}=0.1$ (rightmost curves; see legend).
		}    %/Users/shmuel/Dropbox/Research/Matlab_projects/H2_phases/analytic_approximations/CR_heating_vs_PE_heating
		\label{fig: CR_vs_PE}
\end{figure}

\subsection{PE versus CR}
\label{sub: PE versus CR}

Near solar metallicity ($Z'\sim 1$) photoelectric (PE) heating usually dominates, and the PE heating rate decreases in proportion to the dust-to-gas ratio $Z_d'$.  At sufficiently low $Z'$ heating by cosmic-ray ionization becomes competitive and dominates.
In Fig.~\ref{fig: CR_vs_PE} we show the boundary lines at which the PE and CR heating rates are just equal, in the $I_{\rm UV}/\zeta_{-16}$ versus $Z'$ plane, for $n/\zeta_{-16}$ ratios equal to 0.1, 1, and 10 cm$^{-3}$.
For  $\zeta_{-16}/I_{\rm UV} =1$ the transition from PE to CR heating occurs between $Z'=0.3$ and 0.1, depending on density and temperature.
With decreasing $\zeta_{-16}/I_{\rm UV}$, the transition metallicity is lowered, but not by much.
This is because below $Z'=0.2$ the dust-to-gas ratio falls super-linearly with metallicity (see Eq.~\ref{eq: Zd-Z}).
Conversely, for sufficiently large  $\zeta_{-16}/I_{\rm UV}$ ($\gtrsim 100$), CR heating may dominate even at solar metallicity.

\subsection{Step-1: Reducing the Metallicity}
\label{sub: H2 off}

%The dominant cooling mechanisms for the solar metallicity model were 
%Ly$\alpha$ and \cii+\oi cooling.
%For low metallicity gas, one would expect that H$_2$ cooling would start playing a primary role, because as metallicity decreases the carbon and oxygen elemental abundances decrease.
%On the other hand, at sufficiently low metallicity, H$_2$ formation is dominated by gas-phase reactions (rather than dust formation), and the H$_2$ abundance 
%is then independent of $Z'$.
%At sufficiently low metallicity, H$_2$ cooling must then dominate over \cii and \oi cooling.

We now consider the thermal structures at reduced metallicity, with the exclusion of the H$_2$ heating and cooling processes in {\it Step-1}. Results for $T(n)$ and $P(n)$ are shown as the dashed curves in Fig.~\ref{fig: Tn_diff_Z} 
(top and middle panels)
%and H$_2$ fractions $x_{{\rm H_2}}(n)$ (bottom), 
for $Z'$ from 3 to $10^{-5}$. 
In these computations we have assumed 
$I_{\rm UV}=\zeta_{-16}=1$.

Fig.~\ref{fig: Tn_diff_Z} shows that as $Z'$ is lowered to $0.1$, ($P_{\rm max}$,$P_{\rm min}$) drop to $(4.6,0.43) \times 10^{3}$ cm$^{-3}$ K. These drops are due to the more rapid reduction in $Z'_d$ and associated PE heating rate, compared to the diminished metal line cooling rate that decreases with $Z'$.  
The reduced heating relative to cooling  at $Z'=0.1$, 
reduces the minimum CNM density by a factor of 2.6 (compared to $Z'=1$), to $n_{\rm min}=2.8$ cm$^{-3}$. 
The maximal density of the WNM is also lower, with $n_{\rm max}=0.71$ cm$^{-3}$. 
As the metallicty is reduced further CR heating takes over, and the gas then remains warm to ever higher densities. This is expected given the reduction in the cooling efficiencies for a fixed CR heating rate that is independent of metallicity. Importantly, multiphase behavior is maintained, with $P_{\rm min}$ and $P_{\rm max}$ shifted to higher pressures.

The behavior in the CR heating regime may be analyzed straightforwardly as follows.
As we discussed in \S \ref{sec: Z=1}, as the density is increased in the warm phase, metal line-emission energy losses start contributing at a temperature of $\sim 6\times 10^3$~K. The cooling point temperature is nearly independent of $Z'$, and we denote the corresponding density as $n_{{\rm cool},Z}$ 
\footnote{Here $n_{{\rm cool},Z}$ is the transition point for Ly$\alpha$-to-metal cooling. Later we also introduce $n_{\rm cool, H_2}$ for the transition point from Ly$\alpha$-to-H$_2$ cooling.}. 
For $n>n_{{\rm cool},Z}$, the temperature drops sharply as Ly$\alpha$ becomes inefficient and metal cooling dominates.
%\footnote{we use the subscripts $c_z$ and $c_{\rm H2}$ to denote the point at which metal-line cooling and H$_2$ cooling kick in, resulting in a temperature decrease below $\approx 6 \times 10^3$~K.}.
To obtain an analytic formula for $n_{{\rm cool},Z}$ we equate the heating and metal line cooling rates at $T_0\equiv 6\times 10^3$~K. We assume heating by CR ionization.
This gives
\begin{equation}
\label{eq: point of LyA=OI low Z}
n_{{\rm cool},Z} = \frac{\zeta_{\rm p} E_{\rm cr}}{A_{\rm O} \mathcal{L}_{Z}(T_0) Z'} \
 \approx 4.0 \left( \frac{10^{-2}}{Z'} \right) \zeta_{-16} \ {\rm cm^{-3}} \ ,
\end{equation}
where we have defined the Solar metallicity fine-structure line cooling efficiency (erg s$^{-1}$ cm$^3$) per oxygen atom $\mathcal{L}_Z \equiv \mathcal{L}_{\rm [OI]} +  (x_{\rm C^+}/x_{\rm O}) \mathcal{L}_{\rm [CII]}$ \footnote{As shown in Figs.~\ref{fig: Tn_Z1} and \ref{fig: Cool_heat_diff_Z}, [CII] always dominates over [CI] cooling.
}.
In Eq.~(\ref{eq: point of LyA=OI low Z}) $\zeta_p$ is the primary ionization rate, and $E_{\rm cr}$ is the mean deposition energy per ionization (see Eq.~\ref{eq: CR heating}).
In the numerical evaluation we used $\mathcal{L}_Z(T_0)= 7.2 \times 10^{-23}$
~erg~s$^{-1}$~cm$^3$, 
$x_{{\rm C^+}}/x_{\rm O}=A_{\rm C}/A_{\rm O}=0.56$, $A_{\rm O}=5.4 \times 10^{-4}$ (assuming no depletion), and $E_{\rm cr}=15.2$~eV, $\phi_s=0.57$. 
These values are appropriate for $x_{\rm e} = 8.2 \times 10^{-3}$, the equilibrium electron fraction at the transition point at $Z'=0.01$.
% shmuel7: All the values above, L_Z, E_cr, phi_s, depend on xe which in turn depend back on (zeta/n, T). So to really get these values one needs to either iterate or to write a more complicated equation that takes this secondary dependence into accout. I didn't feel as getting into all these details here, so just wrote the values I adopted.
% \footnote{
% Generally, $\phi_{s}$ and $E_{\rm cr}$ depend on $n/\zeta$ through the electron fraction, however this dependence is weak. }
For $Z'=10^{-2}$ 
Eq.~(\ref{eq: point of LyA=OI low Z}) gives, $n_{{\rm cool},Z}=4.0$~cm$^{-3}$, decreasing linearly with $Z'$, in good agreement with the temperature drop points shown for $T(n)$ in Fig.~\ref{fig: Tn_diff_Z} (dashed curves). For $Z'\gtrsim 0.1$ PE heating contributes in addition to CR, and the cooling densities are larger than given by Eq.~(\ref{eq: point of LyA=OI low Z}).

When the H$_2$ processes are neglected (or negligible) the sharp drop in temperature induced by the transition from Ly$\alpha$ to metal line cooling gives rise to the multiphased $P(n)$ curves, and then $n_{{\rm cool},Z}$ is approximately equal to $n_{\rm max}$ the maximum density for which a WNM is possible. Thus,
\begin{equation}
\label{eq: point of LyA=OI low Z P}
P_{\rm max} \simeq 1.1 n_{{\rm cool},Z} T_0 \
 = 2.6 \times 10^4 \left( \frac{10^{-2}}{Z'} \right) \zeta_{-16} \ {\rm cm^{-3} \ K} \ . 
\end{equation}
{We assume a cosmological helium abundance, $A_{\rm He}=0.1$.
Eq.~(\ref{eq: point of LyA=OI low Z P}) is consistent with our numerical results in Fig.~\ref{fig: Tn_diff_Z}.

	%Eqs.~(\ref{eq: point of LyA=OI low Z}) and (\ref{eq: point of LyA=OI low Z P}) are in good agreement with the numerical results in Fig.~\ref{fig: Tn_diff_Z}, for the low $Z'< 0.1$ (dashed) curves. 
%At $Z'=0.1$ PE heating is still contributing to heating and therefore the transitional density and pressure are higher than those predicted by Eqs.~(\ref{eq: point of LyA=OI low Z}) and (\ref{eq: point of LyA=OI low Z P}).
%Past the transition, the cooling is \cii and \oi  dominated and the equilibrium temperature is given by the solution to
%\begin{align}
%\label{n-T curve G_cr L_Z}
%\frac{\zeta_p E_{\rm cr}}{A_{\rm %O} \mathcal{L}_{z}(T) n Z'} = %1  \ .
%\end{align}
%Thus, at any given temperature, the critical cooling density scales inversely with metallicity, as is also evident in Fig.~\ref{fig: Tn_diff_Z}.
%The consequence is that the minimum and maximum density and pressure at which the gas is multiphased scale as $1/Z'$.
The scalings given by Eqs.~(\ref{eq: point of LyA=OI low Z}) and (\ref{eq: point of LyA=OI low Z P}) are in agreement with previous models \citep[][\citetalias{Wolfire2003}]{Wolfire1995,Liszt2002}.
As we now discuss, when H$_2$ is included this simple picture is greatly altered.

\begin{figure}[t]
	\centering
	\includegraphics[width=0.5\textwidth]{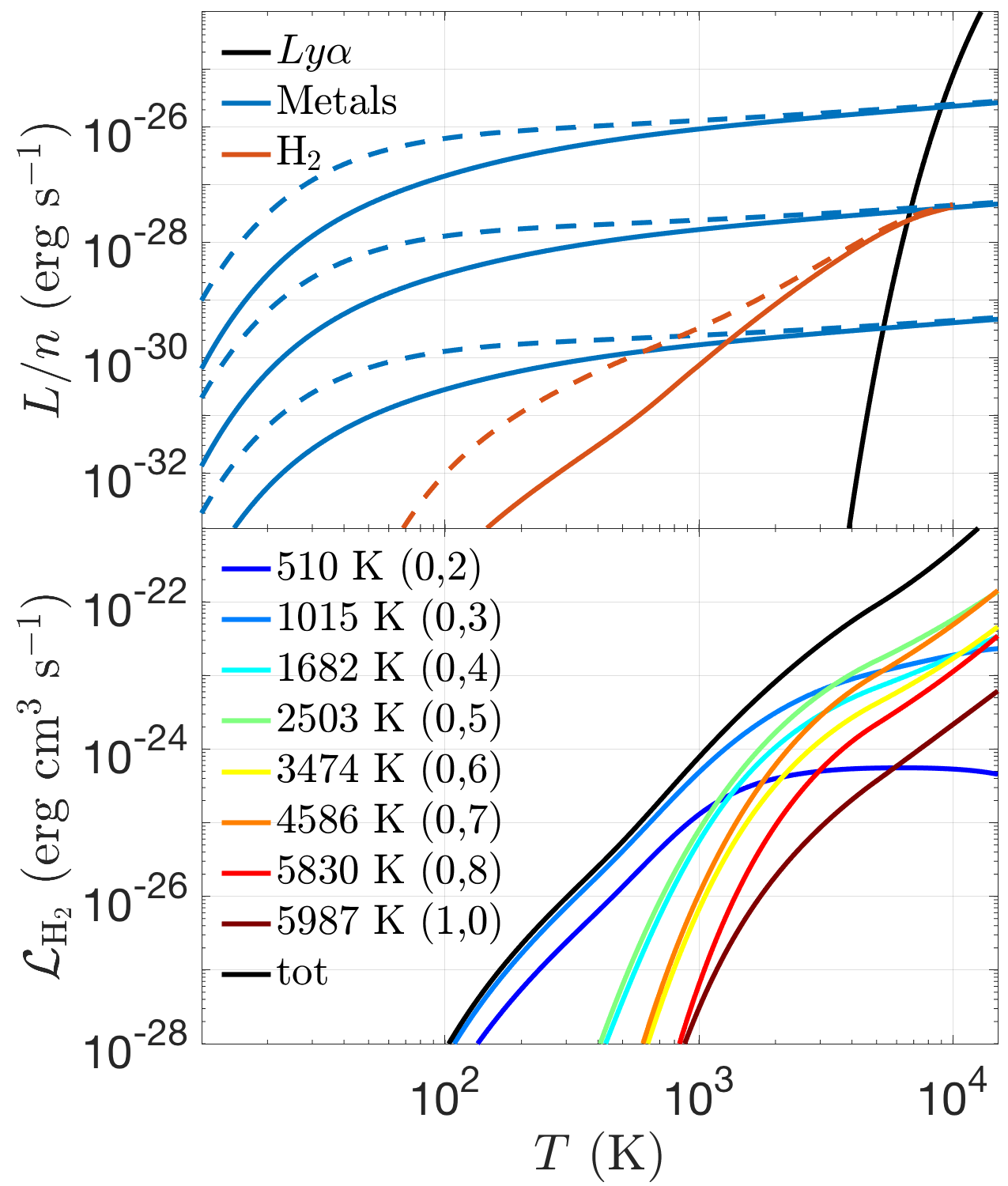} 
	\caption{
Top: The various components of the cooling function per hydrogen nucleus, $L/n$ (erg s$^{-3}$), Ly$\alpha$, metal ([CII], [OI]), and H$_2$ cooling, as functions of temperature. 
We assume $n=1$ cm$^{-3}$, $x_{\rm H^+}=x_{\rm e}=10^{-4}$ (solid) or $10^{-2}$ (dashed), and $x_{\rm H_2}=10^{-6}$.
Typically, for $T \ll 10^4$ K the density is high and $x_{\rm e}$ is small.
For Ly$\alpha$ cooling we always assume $x_{\rm e} =10^{-2}$.
The three sets of metal cooling curves are for $Z'=1, 10^{-2}$, and $10^{-4}$. 
Bottom: The H$_2$ cooling  efficiency $\mathcal{L}_{\rm H_2}$ (erg cm$^3$ s$^{-3}$) by atomic hydrogen excitation (based on data from \citealt{Lique2015}). 
Recall, for low densities, $L_{\rm H_2}=x_{\rm H_2} n^2 \mathcal{L}_{\rm H_2}$.
The contributions from decays of individual ro-vibrationnal H$_2$ levels, $(v,J)$, and their energies above ground (in Kelvin) are indicated.
% The  functions per hydrogen nucleus, as functions of the gas temperature, for $n=1$ cm$^{-3}$, and $x_{\rm H_2}=10^{-6}$.
% The solid and dashed curves are for $x_{\rm e}=x_p=10^{-2}$ and $10^{-4}$, respectively.
% The three sets of metal-line curves correspond to metallicities $Z' = 1$, 10$^{-2}$, 10$^{-4}$, assuming no dust depletion. 
% The H$_2$ curve scales linearly with the assumed H$_2$ abundance. 
% All cooling functions are independent of density in the high density limit, and scale as $n$ in the low density limit.
% Bottom: The relative contribution of  different H$_2$ ro-vibrational levels to the H$_2$ cooling. Excitation by H collisions, and an Ortho-to-para ratio of 3 were assumed.
% The effective temperature of the upper level (in Kelvins), and its vibrational and rotational quantum numbers, $(v,J)$, are indicated.
		}    %/Users/shmuel/Dropbox/Research/Matlab_projects/H2_phases/plots_for_paper/fig0_plot_cool_func_vs_T_diff_Z
		\label{fig: cooling functions}
\end{figure}

\begin{figure}[t]
	\centering
	\includegraphics[width=0.5\textwidth]{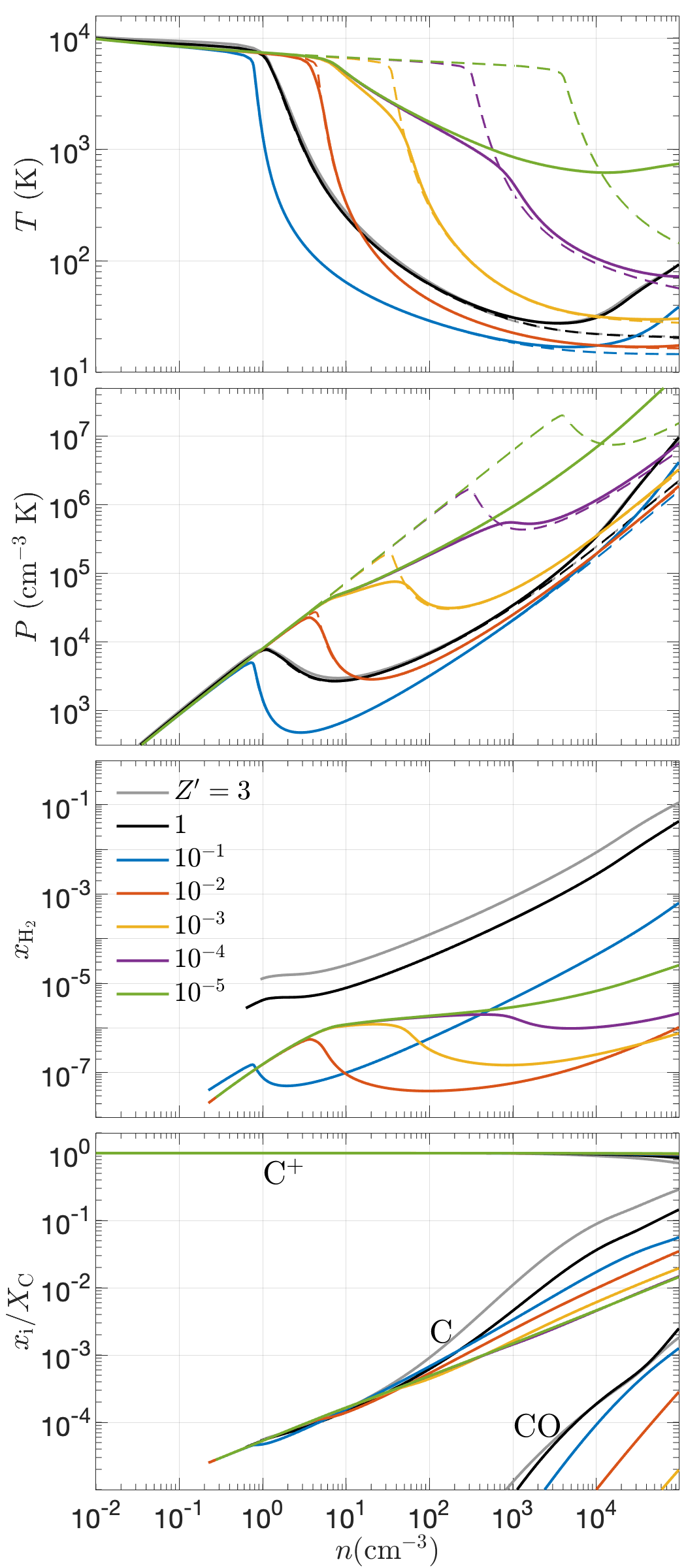} 
	\caption{
Temperature, thermal pressure, H$_2$ abundance, 
and the C, C$^+$ and CO abundances relative to carbon nuclei abundance,
as functions of  the gas hydrogen density, for various metallicity values, from $Z'=3$ down to 10$^{-5}$.
The dashed curves in the top panels are for models that exclude H$_2$ cooling and heating processes.}
% To improve computational speed we did not run the chemica $x_{\rm H_2}$ when $T>8000$ K, since in such conditions Ly$\alpha$ dominates the cooling.}
		\label{fig: Tn_diff_Z}
\end{figure}

\subsection{Step-2: Inclusion of H$_2$ Processes}
\label{sub: H2 ON}

%  Fig.~\ref{fig: T_2d} shows $T$ (left) and $n$ (right) in the two dimentional space of $P$ and $Z'$.

% In the left-hand panel, the cyan and magenta filled points show the minimum allowed $(n, T)$ for the WNM and CNM, under multiphase conditions, and correspond to $P_{\rm min}$ (similarly to the notation used Fig.~\ref{fig: Tn_Z1}).
% The cyan and magenta open points correspond to $P_{\rm max}$.
% In the right-panel, the temperature is shown as a function of $P$, and the multicolored strip corresponds to the multiphase region.
% As the metallicity decreases, the multiphase region becomes smaller, both in the density span and pressure span, until it totally disappears at $Z' \lesssim 2 \times 10^{-4}$.

%  The white curves in  Fig.~\ref{fig: T_2d} (left panel) denote the boundaries at which different cooling processes becomes dominant, 
%  Ly$\alpha$ cooling, H$_2$ cooling, and metal line cooling.
%  These curves intersect at $(Z', n) \sim  (0.01, 2 \ {\rm cm^{-3}})$.  
%   The black curves denote the boundaries beyond which different heating processes becomes dominant, 
% PE heating, CR heating, and H$_2$ heating.
%  These curves intersect at $\approx (Z', n) \sim  (0.1, 10^3 \ {\rm cm^{-3}})$.
%  Both Figs.~\ref{fig: Tn_diff_Z} and \ref{fig: T_2d} assume the fiducial values $I_{\rm UV}=1, \zeta_{-16}=1, N_{s, 19}=1$.
%  We discuss the effect of variation to these parameters in \ref{sub: var UV zeta}.
 
The red curves in the upper panel of Fig.~\ref{fig: cooling functions} show the H$_2$ cooling function.
% as a function of $T$, assuming $n=1$ cm$^{-3}$, $x_{\rm H_2}=10^{-6}$, and $x_{\rm e}=10^{-4}$ (solid) and 10$^{-2}$ (dashed).
In the lower panel we show the contribution of individual
H$_2$ rotational and vibrational levels to H$_2$ cooling, for excitation by atomic hydrogen.
As we discuss below, the fact that (unlike metal cooling) the H$_2$ cooling function never saturates with temperature all the way until Ly$\alpha$ becomes operational, smooths out of the multiphase phenomena and eventually eliminates it entirely at low metallicity.

The solid curves in Fig.~\ref{fig: Tn_diff_Z} show our results for the equilibrium $T(n)$ and $P(n)$ with the H$_2$ processes included in our {\it Step-2}. Again, these are heating via H$_2$ formation, photodissociation, and UV-pumping, and cooling via ro-vibrational line emissions, and collisional dissociation.
The various heating and cooling rates for $Z'=10^{-1,-2...,-5}$, are shown in Fig.~\ref{fig: Cool_heat_diff_Z}.

The efficiencies of the molecular processes depend on the H$_2$ abundance, $x_{\rm H_2}$, that we compute self-consistently with the temperature, for a given gas density and metallicity. In Fig.~\ref{fig: Tn_diff_Z} we show $x_{\rm H_2}$ versus $n$, for $Z'$ from 3 to 10$^{-5}$. 
At all metallicities, H$_2$ destruction by photodissociation is very efficient and the gas remains predominantly atomic.
The H$_2$ abundance is higher at high metallicities as the H$_2$ is formed efficiently on dust-grains.
For $Z'\lesssim 0.1$, the H$_2$ is formed much less efficiently in the gas phase. 
The resulting H$_2$ abundances then range from $10^{-7}$ to $10^{-5}$, consistent with Eq.~(\ref{eq: xH2 gas}).
Interestingly, it is in the low metallicity limit where $x_{\rm H_2}$ is very low, that H$_2$ plays the most significant role in heating and cooling the gas.
At high $Z'$ molecular hydrogen is more abundant but metal line cooling is very efficient and induces a WNM-to-CNM transition at densities well below the densities at which H$_2$ cooling starts playing a role. However, at sufficiently low $Z'$, H$_2$ cooling must eventually dominate over metals.
% at all densities.

% {\color{red} The lower panel of Fig.~\ref{fig: Tn_diff_Z} shows the abundances of C, C$^+$ and CO for the various models.} The carbon is dominated by C$^+$ throughout the parameter space.
% The C and CO abundances increase with $n$, but even at the highest considered density, they reach $\sim 1-10$ \% and $\lesssim 0.1$ \% 
% The oxygen  (not shown in the figure) remains mainly in atomic form.
% As evident from Figs.~\ref{\ref{fig: Tn_Z1} and \ref{fig: Cool_heat_diff_Z}, 

\subsubsection{The onset of H$_2$ cooling at low metallicity}
When the H$_2$ processes are included the $T(n)$ and $P(n)$ curves in Fig.~\ref{fig: Tn_diff_Z} are altered in several ways. First, at sufficiently low metallicity it is H$_2$ rather than the metals that compete with Ly$\alpha$ at the ``cooling point" temperature of $\sim6\times 10^3$~K below which Ly$\alpha$ becomes inefficient.
%shmue6: new
This is also illustrated in the left panels of Fig.~\ref{fig: Cool_heat_diff_Z}.
We define $Z'_{\rm cool,H_2}$ as the critical metallicity at which this switch from metals to H$_2$ occurs at the cooling point. Numerically we find $Z'_{\rm cool,H_2}\approx 10^{-2}$. 
The cooling point density is then shifted to a low value of $\approx 5$~cm$^{-3}$, independent of metallicity (see Figs.~\ref{fig: Tn_diff_Z} and \ref{fig: Cool_heat_diff_Z}).
The H$_2$ abundance at this point is $x_{\rm H_2} \approx 10^{-6}$.

For an analytic estimate we define the gas density for the H$_2$ cooling point where
\begin{align}
\label{eq: point of LyA=H2, 0}
n \mathcal{L}_{\rm H_2} x_{\rm H_2} = \zeta_p E_{\rm cr}\ \ \ .
\end{align} 
We again assume heating by cosmic-ray ionization and $\mathcal{L}_{\rm H_2}$ is the H$_2$ cooling efficiency (erg s$^{-1}$ cm$^3$).
% , and we include the factor of 2 because Ly$\alpha$ still contributes (about equally) at the cooling point\footnote{
% We do not include such a correction in Eq.~\ref{eq: point of LyA=OI low Z} above since unlike the Ly$\alpha$-to-H$_2$ transition, the Ly$\alpha$-to-metal transition is very sharp.
% }. 
With Eq.~(\ref{eq: xH2 gas}) for $x_{\rm H_2}$ with $\eta=1$, we obtain
\begin{equation}
\label{eq: point of LyA=H2}
n_{\rm cool,H_2} = \left( \sqrt{\frac{\alpha_B}{\zeta}} \frac{D_0 I_{\rm UV}}{k_{\ref{reac: H- formation}}} \ \frac{E_{\rm cr} \zeta_p}{\mathcal{L}_{\rm H_2}} \right)^{2/3} 
\approx 5.0 \ \zeta_{-16}^{1/3} I_{\rm UV}^{2/3} \ \ \ {\rm cm^{-3}} \ \ .
\end{equation}
Plugging back into Eq.~(\ref{eq: xH2 gas}) we get the H$_2$ abundance at the cooling point,
\begin{equation}
\label{eq: xH2_cool}
x_{\rm H_2, cool} = 1.4 \times 10^{-6} \left( \frac{\zeta_{-16}}{I_{\rm UV}} \right)^{2/3} \ .
\end{equation}
In the numerical evaluations in Eqs.~(\ref{eq: point of LyA=H2} and \ref{eq: xH2_cool}) we use 
%$\alpha_B=3.7 \times 10^{-13}$~cm$^3$~s$^{-1}$, $k_-=2.3 \times 10^{-15}$~cm$^3$~s$^{-1}$, 
$\mathcal{L}_{\rm H_2} = 2.1 \times 10^{-22}$~erg cm$^3$ s$^{-1}$ at $T_0$, and $E_{\rm cr}$ and $\phi$, as in Eq.~(\ref{eq: point of LyA=OI low Z}).
The corresponding pressure is
\begin{equation}
\label{eq: point of LyA=H2, P}
P_{\rm cool,{\rm H_2}} \approx 1.1 n_{\rm cool,H_2} T_0 \approx 3.3 \times 10^4 \ \zeta_{-16}^{1/3} I_{\rm UV}^{2/3} \ {\rm cm^{-3} \ K} \ \ \ .
\end{equation}
Unlike the Ly$\alpha$-to-metal transition which is accompanied by a maximum in the $P-n$ curve (Eq.~\ref{eq: point of LyA=OI low Z P}), the Ly$\alpha$-to-H$_2$ transition does not induce such a maximum but only a kink in the $P-n$ curve (see Fig.~\ref{fig: Tn_diff_Z}).

% We obtain an analytic estimate for the critical metallicity at the cooling point by equating the metal and H$_2$ cooling rates and evaluating at $T_0=6\times 10^3$~K. This is equivalent to equating Eq.~(\ref{eq: point of LyA=OI low Z}) and (\ref{eq: point of LyA=H2}) for the cooling densities. This gives
The critical metallicity below which H$_2$ cooling becomes important
is
\begin{equation}
\label{eq: Z_c_H2}
Z'_{\rm cool,H_2} \equiv \frac{\mathcal{L}_{\rm H_2} x_{\rm H_2}}{\mathcal{L}_{z} A_{\rm O}} \approx 8.0 \times 10^{-3} \left( \frac{\zeta_{-16}}{I_{\rm UV}} \right)^{2/3} \ .
\end{equation}
As we discuss in \S \ref{sec: Parameter Study}, this 
formula is in good agreement with our numerical results. 
Importantly, the temperature decreases more gradually for transitions from Ly$\alpha$ to H$_2$ cooling. Because of the contributions from many lines in the ro-vibrational ladder (as illustrated in the bottom panel of Fig.~\ref{fig: cooling functions}), the H$_2$ cooling function remains sensitive to temperature in the entire range $10^2< T < 10^4$~K.
This is illustrated in the upper panel of Fig.~\ref{fig: cooling functions} where the red curves show the much steeper H$_2$ cooling function compared to the flat metals function (blue curves). For this reason, the temperature drop at $n_{\rm cool,H_2}$ is moderated and does not produce a local maximum in the $P(n)$ curve. Most generally, {\it the transition from Ly$\alpha$ to H$_2$ cooling does not result in a multiphased behavior.}

\begin{figure*}[t]
	\centering
	\includegraphics[width=1\textwidth]{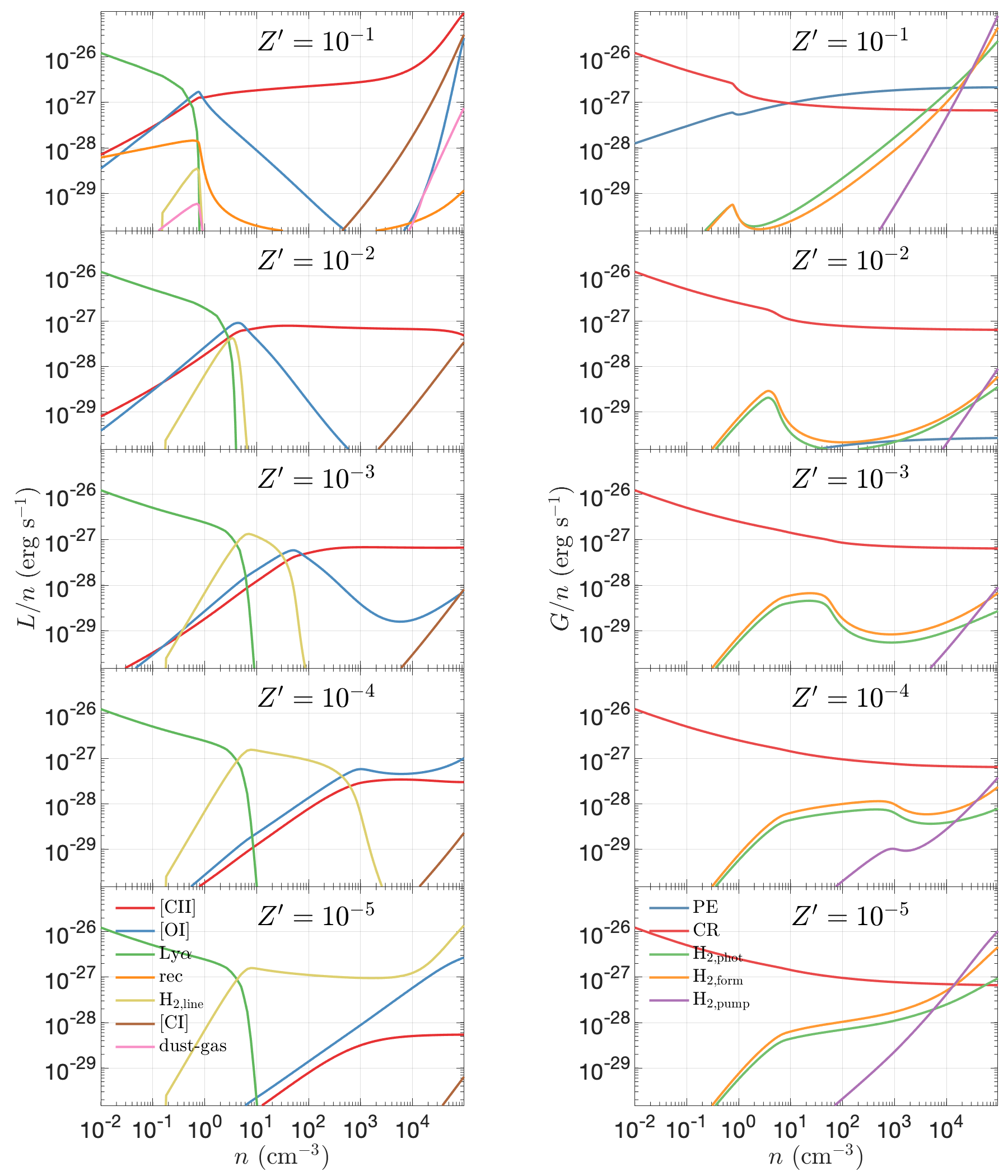} 
	\caption{
Cooling (left) and heating (right) rates per hydrogen nucleus (erg s$^{-1}$), as functions of  the gas hydrogen density, at various metallicities.
}  %Heat_cool_diff_Z.m
		\label{fig: Cool_heat_diff_Z}
\end{figure*}

While H$_2$ cooling does not induce a multiphase, Fig.~\ref{fig: Tn_diff_Z} shows that multiphased behavior continues to be produced by metal-line cooling, although over a much narrower pressure range, down to a metallicity as low as $10^{-4}$. 
This multiphase is induced by the transition from H$_2$ to metal cooling (rather than Ly$\alpha$ to metals) and occurs for a lower cooling point temperature of $T_0\approx 600$ K. So long as the heating is by CR ionization the cooling point density at which the sharp temperature drop occurs is still given by Eq.~(\ref{eq: point of LyA=OI low Z}) but evaluated at the lowered $T_0$. Thus, in this regime $n_{{\rm cool},Z}$ continues to vary as $1/Z'$, but the density prefactor in Eq.~(\ref{eq: point of LyA=OI low Z}) is reduced.
For example, for $Z'=10^{-3}$, $n_{{\rm cool},Z}=40$ cm$^{-3}$, leading to a pressure maximum of $P_{\rm max}\approx 1.1 n_{{\rm cool},Z}T_0=7.6 \times 10^{4}$ cm$^{-3}$~K for the WNM (see Fig.~\ref{fig: Tn_diff_Z}). 

Because H$_2$ cooling reduces the temperature of the WNM, the multiphase strip is now narrower than at solar metallicity, and continues to shrink with decreasing metallicity.
For example, while at solar metallicity the multiphase range is wide, with $\delta_n \equiv (n_{\rm max}-n_{\rm min})/n_{\rm min}=5.4$ and $\delta_P \equiv (P_{\rm max}-P_{\rm min})/P_{\rm min}=1.9$, 
for $Z'=10^{-3}$ we have $\delta_n=3.8$ and $\delta_P=0.88$.
For $Z'=10^{-4}$, the multiphase phenomena is already extremely suppressed, with $\delta_n=1.6$ and $\delta_P=0.05$.

\subsubsection{The onset of H$_2$ heating at extremely low metallicity}
For still lower metallicities ($Z' \leq 10^{-5}$) sharp gas cooling and associated multiphased structures disappear entirely.
The density at which metal cooling can balance CR heating increases as $1/Z'$,
% (since metal cooling is $\propto Z' n^2$ while CR heating is $\propto n$).
and at sufficiently low metallicities this density becomes so high that H$_2$ heating becomes operational and dominant. In this limit metal cooling becomes ineffective. This is  because H$_2$ heating includes a dependence on the molecular formation rate (see Eq.~\ref{eq: H2 heating}) so that the heating rate increases more rapidly than $n^2$. However, metal cooling varies no faster than $n^2$, and so cannot compete. In this limit H$_2$ is the dominant coolant, multiphase behavior does not occur, and the gas is driven to 
intermediate, thermally stable, temperatures of $\sim 600$~K.
%shmuel6: new
This behavior is illustrated in Figs.~\ref{fig: Tn_diff_Z} and \ref{fig: Cool_heat_diff_Z}. 
% While at $Z' \geq 10^{-4}$, there always occur a transition from H$_2$ cooling to metal cooling (or if the metallicity is sufficiently high, from Ly$\alpha$ to metal cooling).
% At very low metallicity, the H$_2$ heating mechanisms which dominate at high density introduce a heating rate that
% cannot be balanced by metal cooling, and H$_2$ line emission dominates the cooling for all densities above $n \approx 5$ cm$^{-3}$.

We obtain an analytic estimate for the critical metallicity, $Z'_{\rm heat,H_2}$, below which metal line cooling is no longer effective as follows. We first estimate the density $n_{\rm heat,H_2}$ at which H$_2$ and CR heating are equal. For $n>n_{\rm heat,H_2}$ we are in the H$_2$ heating regime. The critical $Z'_{\rm heat,H_2}$ is then the metallicity at which the metal cooling density, $n_{{\rm cool},Z}$, as given by Eq.~(\ref{eq: point of LyA=OI low Z}), is at least as large as $n_{\rm heat,H_2}$.  For $n_{{\rm cool},Z} \gtrsim n_{\rm heat,H_2}$ metal line cooling is ineffective because it cannot compete with H$_2$ heating. 

Equating the H$_2$ and CR heating rates gives,
\begin{align}
\label{eq: n_crit H2 heating}
n_{\rm heat,H_2} &= \left( \frac{ \zeta_p E_{\rm cr} n_{\rm crit}}{E_{\rm H_2,2} k_{\ref{reac: H- formation}} \sqrt{\zeta/\alpha_B}} \right)^{2/3} \\
&= 7.2 \times 10^3 \ \zeta_{-16}^{1/3} \ T_3^{-1.08} \ {\rm cm^{-3}} \ ,
\end{align}
where we used our H$_2$ heating expression, Eq.~(\ref{eq: H2 heating}), assuming UV pumping heating dominates (but still in the $n<n_{\rm crit}$ limit), and used Eq.~(\ref{eq: R gas phase}) for the effective gas-phase formation rate coefficient $R_-$.
% Eq.~(\ref{eq: n_crit H2 heating}) holds for volume densities $n<n_{\rm crit}$, and at low metallicities where H$_2$ forms in gas phase, and PE heating is negligible (roughly below $Z'=0.1$ - see Fig.~\ref{fig: CR_vs_PE}). 
% In the second equality we used $n_{\rm crit}=1.1 \times 10^5/\sqrt{T_3}$ cm$^{-3}$, $E_{\rm eff}=20.7$ eV, and $E_{\rm cr}=8.1$ eV, $\phi_s = 0.67$ (appropriate for $n/\zeta_{-16}=10^3$ cm$^{-3}$).
    Next, 
equating the metal cooling density $n_{{\rm cool},Z}$ (Eq.~\ref{eq: point of LyA=OI low Z}) with $n_{\rm heat,H_2}$ above gives,
\begin{equation}
\label{eq: Z_h_H2}
Z'_{\rm heat,H_2} = 1.2 \times 10^{-5} \  \zeta_{-16}^{2/3} \ .
\end{equation}
In the numerical evaluation we adopted a metal cooling efficiency $\mathcal{L}_z=1.5 \times 10^{-23}$ erg cm$^{3}$ s$^{-1}$ at $n= 10^4$ cm$^{-3}$, $T=10^3$~K.
% at which metal cooling {\it does} take over from H$_2$ cooling when CR heating dominates. 
For $Z' \lesssim Z_{\rm heat,H_2}$
 the gas never undergoes a second metal-line cooling phase, and there is no multiphase structure. 
 This formula is in agreement with the numerical results shown in Fig.~\ref{fig: Tn_diff_Z}.
 
 %shmuel6: new
Eq.~(\ref{eq: Z_h_H2}) is a strict requirement on the metallicity, expressing the metallicity below which a multiphase {\it cannot} possibly occur.
In practice, as discussed above, much before the metallicity becomes that low, the multiphase region may already become very narrow, so that any thermal instability effects may become negligible.

\section{Parameter Study}
\label{sec: Parameter Study}

\begin{figure*}[t]
	\centering
	\includegraphics[width=1\textwidth]{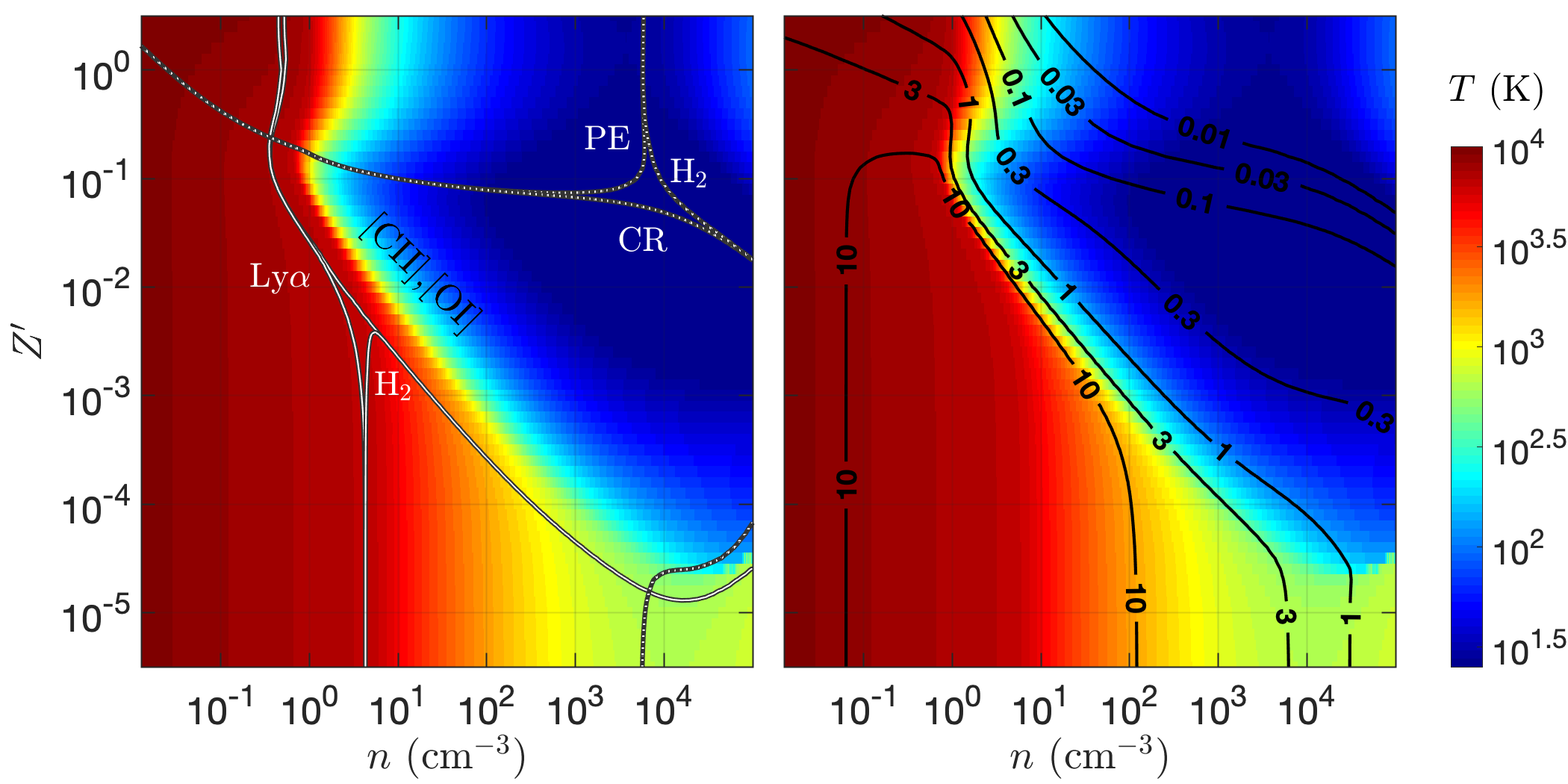} 
		\includegraphics[width=1\textwidth]{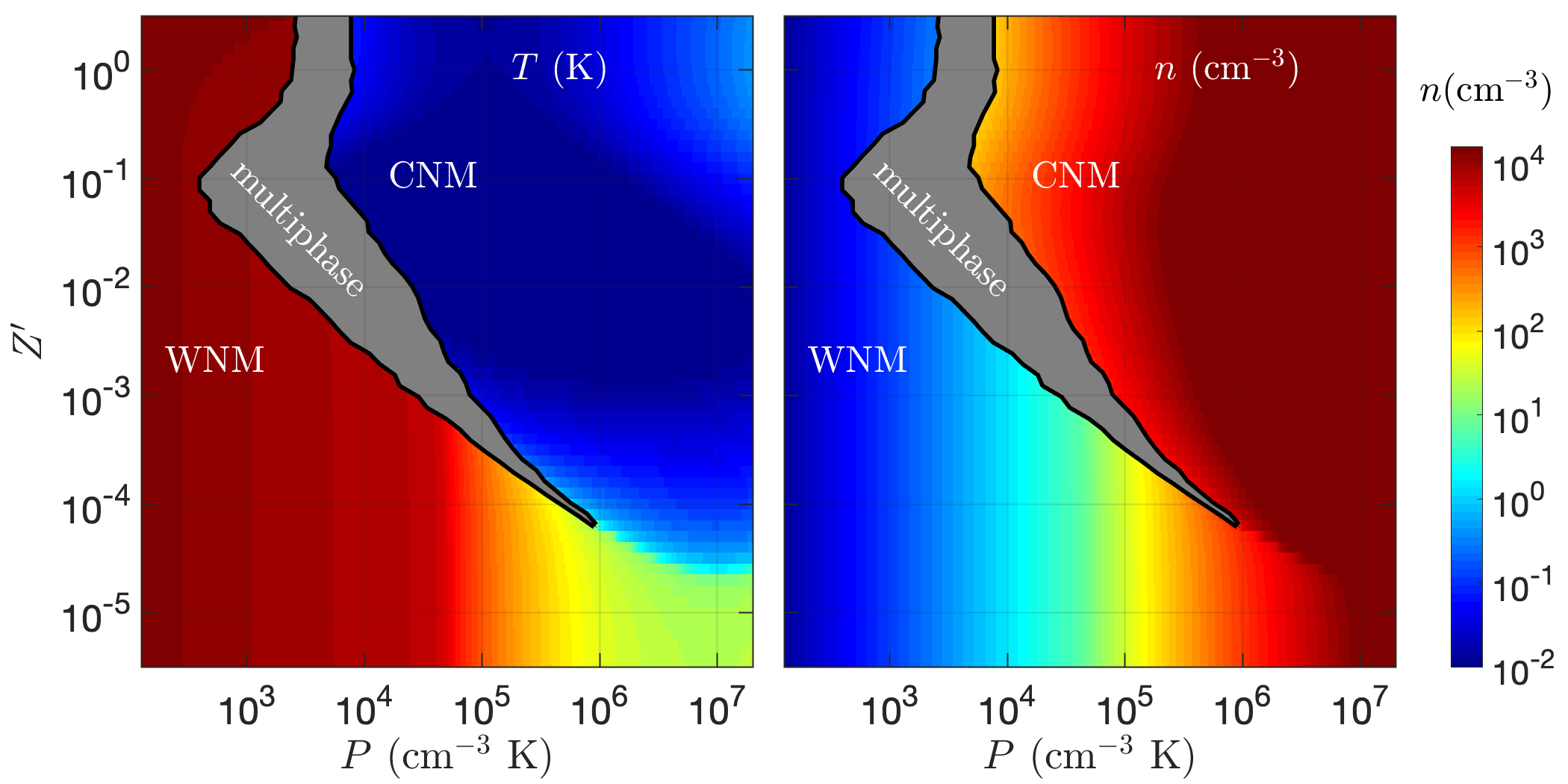} 
	\caption{
Top: The temperature as a functions of metallicity and gas density.
In the left panel, the solid contours delineate the regions where Ly$\alpha$-, metal-, and H$_2$-cooling become dominant. The dashed contours delineate PE-, CR-, and H$_2$-heating dominated regions.
The contours in the right panel are the cooling times in Myrs.
Bottom: The temperature (left) and gas density (right) as a functions of metallicity and pressure.
The WNM, CNM and multiphase regions are indicated.
A 3D interactive surface plot is presented in the online version (see also
\href{https://plot.ly/~sb2580/47}{https://plot.ly/~sb2580/47}).
		}    %/Turb//plots_for_paper/profiles_and_N1_tot_vs_
		\label{fig: T_2d}
\end{figure*}

In this section, in Figs.~\ref{fig: T_2d} and \ref{fig: T_2d_vars}, we present results of our thermal phase computations in the form of equilibrium temperature contour plots in the metallicity versus density ($Z'-n$) plane, as well as temperature and density contour plots in the metallicity versus pressure ($Z'-P$) plane. 
We investigate the effects of variations in $\zeta$ and $I_{\rm UV}$.
We also discuss the time scales relevant for our assumptions of thermal and chemical equilibrium.

\subsection{Contour plots}
\label{sub: contourplots}

The top panels of Fig.~\ref{fig: T_2d} show the equilibrium gas temperature in the metallicity versus density ($Z'-n$) plane, for $\zeta_{-16}=I_{\rm UV}=1$. The colors indicate the temperature, ranging from cold $\sim 10^2$~K (blue), to intermediate $\sim 10^3$~K (yellow), to warm $\sim 10^4$~K (red). 
In the upper-right panel we
overplot contours of the cooling times.
 The cooling times range from $\sim 1-10$ Myrs in the WNM, depending on the metallicity, 
down to 0.1-0.4 Myrs in the CNM at low $Z'$, and down to less then 0.01  Myrs at high $Z'$.
At higher metallicities the increase in the heating rate due to the onset of PE heating, is accompanied by appropriately higher cooling rates. 
For the CNM, the cooling rate naturally increases with increasing metallicity, as \cii and \oi dominates the cooling.
In the WNM, where Ly$\alpha$ dominates, the cooling rate increases via a slight increase in the equilibrium temperature. 
Since at low $Z'$ a large portion of the parameter space is WNM, or the intermediate $\approx 10^3$ K medium (at very low $Z'$), with cooling times 1-10 Myrs, the gas may potentially be out of thermal equilibrium, depending on the magnitude of the local dynamical time (e.g., the sound crossing time, turbulent time, free-fall time, etc.).
We further discuss the cooling and chemical time in the Appendix.

In the upper-left panel the various curves delineate the dominating heating and cooling processes that operate in different regions of parameter space. For heating, these are (a) photoelectric (PE) heating, above the dashed curve from the WNM into the CNM, (b) H$_2$ heating, to the right of the dashed curve at high density in the CNM at moderately reduced metallicity, and again at high densities and very low metallicities in the intermediate temperature zone, and (c) cosmic-ray (CR) heating everywhere else. For cooling these are (a) Ly$\alpha$ to the left of the solid curve in the low density WNM, (b) metals to the right of the solid curve into the CNM, and (c) H$_2$ cooling below the solid curve in the lower right portion of parameter space.

For any $Z'$ the gas temperature decreases monotonically with increasing $n$, with sharp drops in $T$, from the WNM to CNM (red to blue), occurring near the Ly$\alpha$/metal or H$_2$/metal cooling point boundaries. The WNM-to-CNM transition points depend on metallicity. The intermediate temperature zone appears at very low $Z'$ and high density.

For $Z'\gtrsim 0.2$ PE heating dominates over CR ionization in the WNM. Because PE heating and metal cooling both vary linearly with $Z'$ (see Eq.~\ref{eq: Zd-Z}), in this regime the Ly$\alpha$ to metal cooling point runs vertically (independent of $Z'$) at an approximately constant density of $\approx 0.5$~cm$^{-3}$. Below $Z'\approx 0.2$ CR heating takes over, and the cooling point then moves diagonally, with $n_{{\rm cool},Z}\propto 1/Z'$, consistent with Eq.~(\ref{eq: point of LyA=OI low Z P}) for a cooling point temperature $T_0=6\times 10^3$~K appropriate for the transition from Ly$\alpha$ to metal cooling. Below a metallicity of $8\times 10^{-3}$, H$_2$ cooling (rather than metals) takes over from Ly$\alpha$ in the WNM. This is  consistent with our analytic estimate (Eq.~\ref{eq: Z_c_H2}) for $Z'_{\rm cool, H_2}$. The Ly$\alpha$ to H$_2$ boundary line is vertical (independent of $Z'$) and occurs well inside the WNM at a density $\approx 3$~cm$^{-3}$ consistent with our Eq.~\ref{eq: point of LyA=H2} for $n_{\rm cool,H_2}$.
Below $Z'_{\rm cool, H_2}\approx 8\times 10^{-3}$ the WNM-to-CNM boundary remains diagonal with $n_{{\rm cool},Z}\propto 1/Z'$ as given by Eq.~(\ref{eq: point of LyA=OI low Z P}), but with a lower $T_0$ (i.e., $T_0=5000$ to $600$ K, for $\log Z'=-2.3$ to $-4.7$), as appropriate for the transition from H$_2$ to metal cooling.

For $Z'< Z'_{\rm heat,H_2}\approx 10^{-5}$ and $n> n_{\rm heat,H_2} \approx 10^4$~cm$^{-3}$, H$_2$ heating dominates over CR ionization, and therefore metal cooling becomes ineffective.  The critical values $Z'_{\rm heat,H_2}$ and $n_{\rm heat,H_2}$ occur at the intersection of the CR-to-H$_2$ heating and H$_2$-to-metal cooling boundary curves in Fig.~\ref{fig: T_2d}, and are consistent with our analytic estimates given by Eqs.~(\ref{eq: n_crit H2 heating}) and (\ref{eq: Z_h_H2}).
In this portion of parameter space an intermediate equilibrium temperature of $\sim 10^3$~K is set by the balance of H$_2$ heating and cooling.

The lower panels of Fig.~\ref{fig: T_2d} show the temperature (left) and density (right) in the pressure versus metallicity ($P-Z'$) plane (again for $\zeta_{-16}=I_{\rm UV}=1$). The regions of pure CNM (cold/dense) and WNM (warm/diffuse) are indicated, as is the intermediate temperature H$_2$ zone at very low metallicity and high density. The multiphased region where the CNM and WNM overlap occurs between the solid black curves, from $P_{\rm min}$ (to the left) to $P_{\rm max}$ (to the right).  

The position of the multi-phased zone in Fig.~\ref{fig: T_2d} is determined primarily by (a) the switch from PE to CR heating, (b) the dependence of the metal cooling point on $Z'$, and (c) the onset of H$_2$ cooling and heating at low metallicity.
For $Z'$ down to $\sim 0.1$ PE heating dominates, and the multiphased zone widens and occurs at reduced values of $P_{\rm min}$ and $P_{\rm max}$. The width is maximal at $Z' \approx 0.1$ where PE heating is least effective compared to metal cooling. For $Z'\lesssim 0.1$ CR heating takes over (independent of metallicty) and $P_{\rm min}$ and $P_{\rm max}$ increase as the transition to metal cooling requires larger densities. 

As the metallicity is lowered below $Z'_{\rm cool,H_2}\approx 8 \times 10^{-3}$ the multiphased zone become narrower due to H$_2$ cooling in the WNM.
At $Z' \approx 10^{-4}$ there still exists a (small) temperature jump  at the WNM to CNM boundary, but the pressure range of the multiphase zone is already negligible.
At $Z' < Z'_{\rm heat,H_2}\approx 10^{-5}$ metal cooling becomes ineffective, the temperature then decreases smoothly with increasing pressure, from the $\sim10^4$~K Ly$\alpha$ cooled WNM, to the $\sim 10^3$~K H$_2$ cooled intermediate temperature zone.

\subsection{Varying $\zeta$, $I_{\rm UV}$ and $D_-$}
\label{sub: var UV zeta}

\begin{figure*}[t]
	\centering
	\includegraphics[width=1\textwidth]{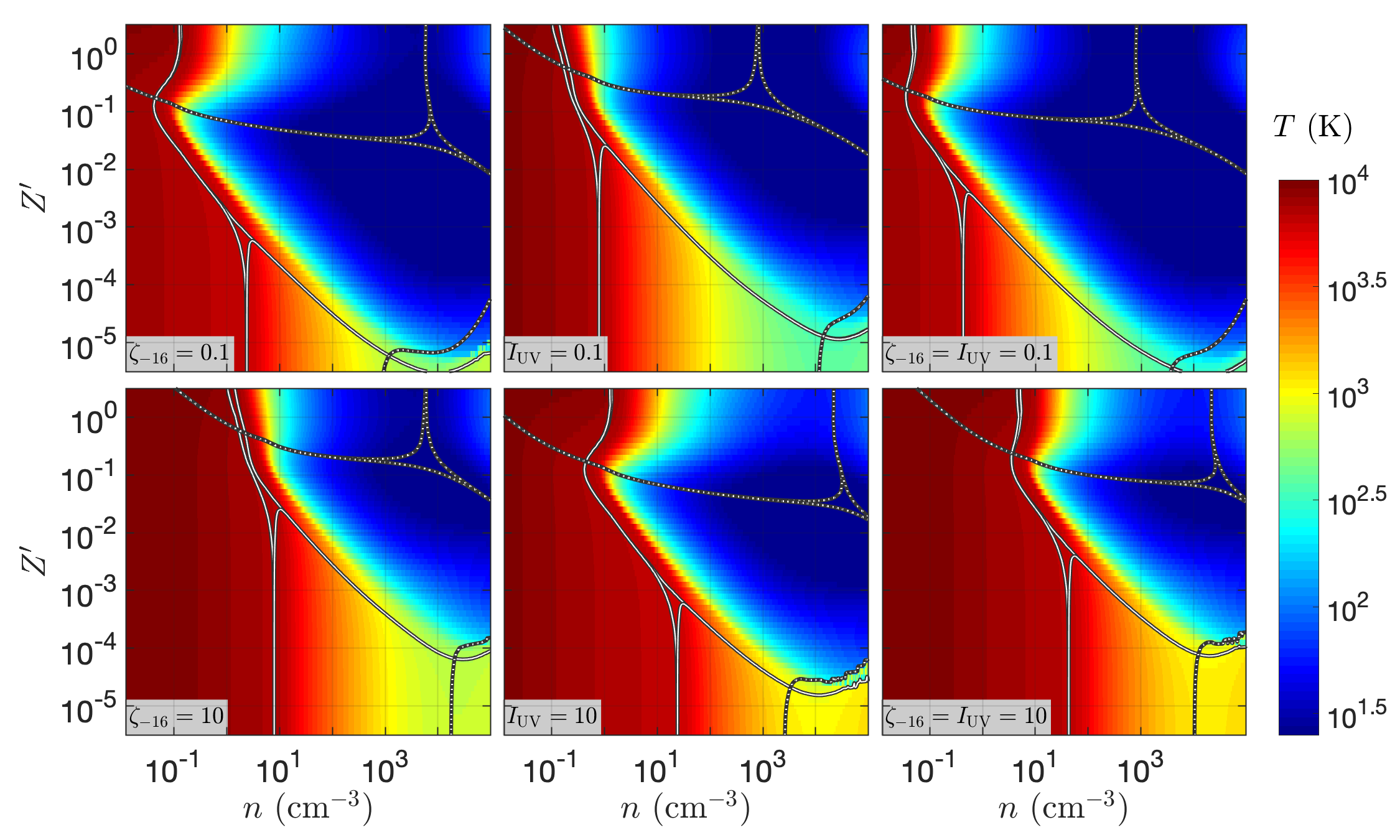} 
		\includegraphics[width=1\textwidth]{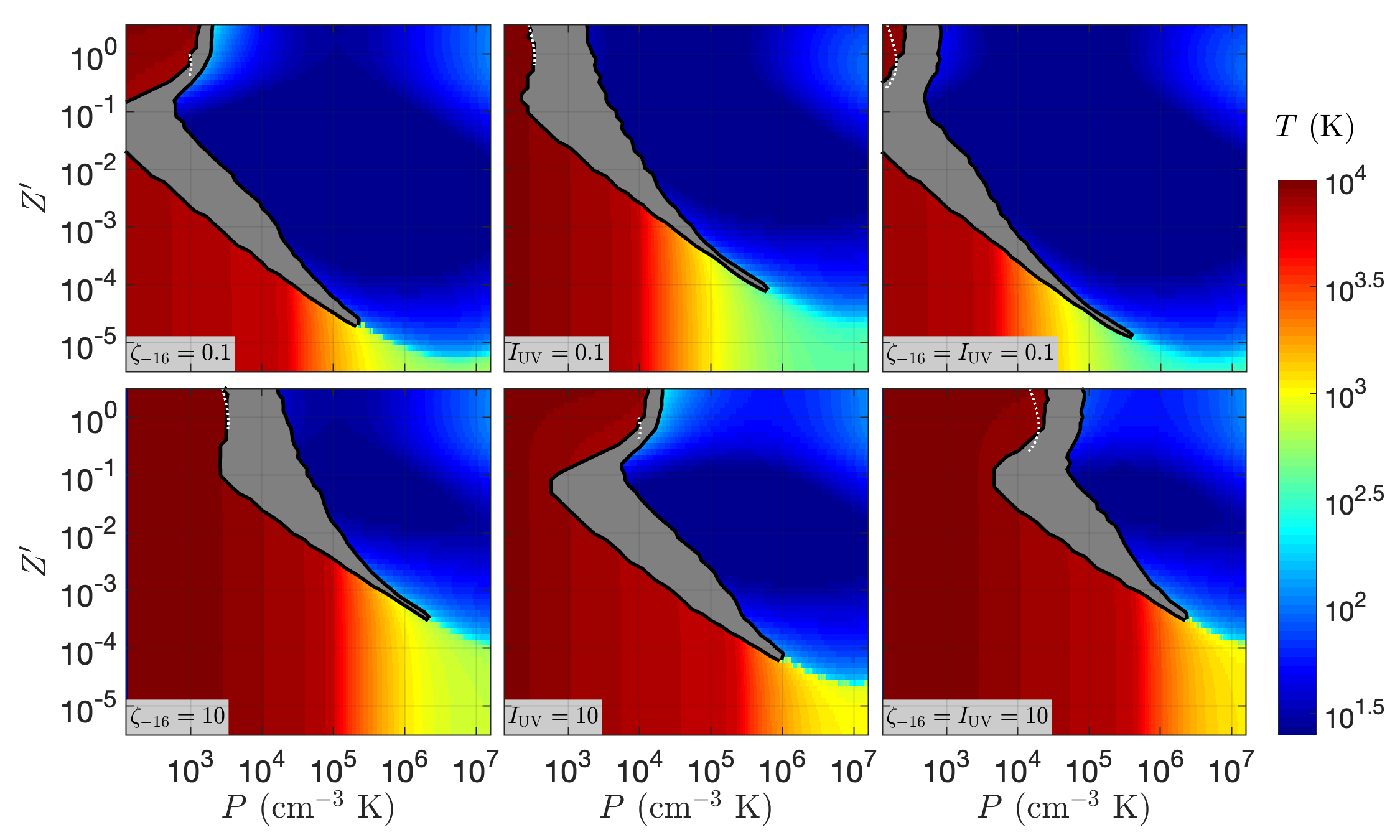} 
	\caption{
As Fig.~\ref{fig: T_2d}, but assuming variations in $\zeta$ and $I_{\rm UV}$.
% Left-row: $\zeta_{-16}=0.1$ and 10, $I_{\rm UV}=1$, middle row: $I_{\rm UV}=0.1$ and 10, $\zeta_{-16}=1$, right-row: $\zeta_{-16}=I_{\rm UV}=0.1$ and 10.
The contours, and shaded regions have the same meaning as in Fig.~\ref{fig: T_2d}.
The dotted curves in the lower panels are the \citetalias{Wolfire2003} formula for $P_{\rm min}$ (Eq.~\ref{eq: Pmin Wolfire}).
		}    %/Turb//plots_for_paper/profiles_and_N1_tot_vs_
				\label{fig: T_2d_vars}
\end{figure*}

In Fig.~\ref{fig: T_2d_vars} we explore the thermal structures for varying UV intensities and cosmic-ray ionization rates.
%The upper (lower) two rows show the temperature in the $n-Z'$ ($P-Z'$) parameter space.
The left column shows the cases $\zeta_{-16}=0.1$ and 10, assuming $I_{\rm UV}=1$.
The middle column is for $I_{\rm UV}=0.1$ and 10, with $\zeta_{-16}=1$.
The right column shows models in which $I_{\rm UV}$ and $\zeta_{-16}$ scale together by $\pm 1$dex relative to our fiducial case $\zeta_{-16}=I_{\rm UV}=1$.
%The middle and right panels of Fig.~\ref{fig: T_2d_vars} show the effect of a varying UV intensity, once with $\zeta$ kept constant (middle-panel), and once with $\zeta_{-16}=I_{\rm UV}$ (right-panel).
%The former case has the advantage that it isolates the effect of varying $I_{\rm UV}$.
%It also illustrates the effect of a varying UV spectrum shape, which would change the H$_2$ dissociation rate (i.e., a 10$^5$ K blackbody, versus a Draine spectrum).
This last set reflects a physical scenario in which the massive stars that produce the energetic UV photons, are also the sources of the cosmic-rays originating in the remnants of core-collapse supernovae.
For the $Z'-n$ plots, the PE/CR/H$_2$ and Ly$\alpha$/metals/H$_2$ heating and cooling boundaries are shown as the solid and dashed curves, as in the upper left panel of Fig.~\ref{fig: T_2d}. For the $Z'-P$ plots, the various phases including the multiphased zones are indicated as in the lower panels of Fig.~\ref{fig: T_2d}.

%AMIRL77 Some adjustments to the text here. In the figure iteself I think you should thicken the dotted curves, they are hard to see now.
In the lower panels we also show, as dotted white curves, a comparison to the \citetalias{Wolfire2003} formula for $P_{\rm min}$ as a function of $I_{\rm UV}$, $Z'$, $Z'_d$, and $\zeta_{-16}$,
\begin{equation}
\label{eq: Pmin Wolfire}
    P_{\rm min} = 8500 I_{\rm UV} \frac{Z_d'/(Z' f)}{1+3.1(I_{\rm UV}Z_d'/\zeta_{-16})^{0.365}} \ {\rm cm^{-3} \ K} \ ,
\end{equation}
(their Eq.~33) over its validity range (their Eqs.~35-38).
We include the factor $f \equiv (1-\delta_{\rm C} Z_d'/Z')/(1-\delta_{\rm C})$ to correct\footnote{While \citetalias{Wolfire2003} use elemental abundances similar to ours for $Z'=1$, they do not 
have a metallicity dependent depletion in their model, and $f$ corrects their formula for this effect.} for the metallicity dependent depletion onto dust (i.e., our Eq.~\ref{eq: C and O abundances}).
For $Z_d'$ we use our Eq.~(\ref{eq: Zd-Z}).
The agreement is not perfect due to the different treatment of cosmic-ray and X-ray heatings, and PAH chemistry.
However, overall, the general trends with $Z'$, $\zeta$ and $I_{\rm UV}$ are recovered.

The left-panels of Fig.~\ref{fig: T_2d_vars} show that as $\zeta$ increases, the WNM, CNM and the multiphase regions, are all shifted to higher densities and pressures. Within the CR-heating region (below the almost horizontal dotted line)
higher densities are required for cooling to offset an increasing CR heating rate.
Within the PE-heating region,
as $\zeta$ increases, the electron abundance increases, the dust-grains are less positively charged, and the release of photoelectrons from the dust-grains into the gas is more efficient (i.e., see Fig.~1 in \citealt{Wolfire1995}).
In the middle panels, where $I_{\rm UV}$ varies (with $\zeta$ kept constant), the shift is observed only at high metallicities ($Z' \gtrsim 0.1$) where PE heating dominates, as expected.

% The diagonal white lines, in all panels, are the Ly$\alpha$-to-metal cooling transition lines.
% Within the CR-heating dominated region, the location of this line
% is well described by Eqs.~(\ref{eq: point of LyA=H2,2} - \ref{eq: point of LyA=OI low Z P}). 
% As this line is always located near the maximum WNM density, Eq.~(\ref{eq: point of LyA=H2,2} - \ref{eq: point of LyA=OI low Z P}) are also useful in determining the WNM-to-CNM transition point.

%As expected, the region dominated by CR heating becomes larger and the PE heating region becomes smaller, when $\zeta$ is increased at constant $I_{\rm UV}$ (left panels), or when $I_{\rm UV}$ is decreased at constant $\zeta$ (middle panels). 
%The white curves in the upper panels are the locus of points at which metal-cooling (diagonal curve) and H$_2$-cooling (vertical curve) kick-in, as indicated in Fig.~\ref{fig: T_2d_vars}.

%In both the CR dominated regime and PE dominated regime w
%Within the UV-heating dominated regime, the Ly$\alpha$-to-metal cooling transition is well

The critical metallicity at which H$_2$ cooling becomes important, $Z'_{\rm cool, H_2}$, increases 
with $\zeta$ and decreases with $I_{\rm UV}$.
When $\zeta$ and $I_{\rm UV}$ scale together, $Z'_{\rm cool, H_2}$ remains unchanged.
This is because the H$_2$ abundance (a) increases with $\zeta$, as the gas phase formation is more efficient, and (b) decreases with $I_{\rm UV}$, as photodissociation is more efficient.  
The observed dependence of $Z'_{\rm cool, H_2}$ on $\zeta$ and $I_{\rm UV}$ is in good agreement with Eq.~(\ref{eq: Z_c_H2}).
The location of the Ly$\alpha$-to-H$_2$ transition (the vertical white contour) is in good agreement with Eq.~(\ref{eq: point of LyA=H2}). 
For example, looking at the left panels of Fig.~\ref{fig: T_2d_vars} we see that as $\zeta_{-16}$ increases from 0.1 to 10 (with $I_{\rm UV}=1$ kept constant), the Ly$\alpha$-to-H$_2$ transition shifts from $n \approx 2$ to 8 cm$^{-3}$.
On the other hand, when $I_{\rm UV}$ varies from 0.1 to 10 (and $\zeta_{-16}=1$ is kept constant, middle panels), the transition point shifts from $\approx 0.7$ to $\approx 20$ cm$^{-3}$.
Finally, when both $\zeta$ and $I_{\rm UV}$ vary together from 0.1 to 10, the transition point shifts by two orders of magnitude, from $\approx 0.4$ to 40  cm$^{-3}$.

The critical metallicity, $Z_{\rm heat, H_2}$, at which H$_2$ heating takes over, increases with $\zeta$, and is independent of $I_{\rm UV}$, as predicted by Eq.~(\ref{eq: Z_h_H2}).
As discussed in \S \ref{sub: H2 ON}, this metallicity is set by 
the requirements that H$_2$ heating equal CR heating, which in turn must equal metal cooling, all of which are independent of $I_{\rm UV}$.
An increase in the CR intensity increases both the H$_2$ formation rate and the CR heating rate, but the latter is affected more strongly, and thus $Z_{\rm heat, H_2}$ increases with $\zeta$.

As discussed in \S \ref{sub: params: IUV}, the H$^-$ photodetachment rate at low metallicity is uncertain. We consider variations between $D_-=5.6 \times 10^{-9}$ s$^{-1}$ for the extrapolated Draine field (our fiducial models), up to $D_-=2.7 \times 10^{-7}$ s$^{-1}$ for a Draine+Mathis field.
To investigate the effect of a varying $D_-$, we have also computed a set of models with the higher Draine+Mathis $D_-$ rate.
The resulting temperature map is identical to that in Fig.~\ref{fig: T_2d}, with the only difference that the H$_2$ cooling line is shifted to a higher density, $n_{\rm cool, H_2}=15$ cm$^{-3}$, i.e., an increase by a factor of 3.5.
Since the $n_{\rm cool,Z}$ curve is unaffected by variations in $D_-$ (as it is set by metal cooling), the critical metallicity for H$_2$ cooling is reduced, and $Z_{\rm cool, H_2}$ is reduced to $10^{-3}$.
The shift to a higher cooling density is expected as the higher detachment rate reduces the H$_2$ formation rate, requiring a higher gas density for the onset of H$_2$ cooling.
Unlike $n_{\rm cool,H_2}$ and $Z_{\rm cool, H_2}$, the heating points $n_{\rm heat,H_2}$ and $Z_{\rm heat,H_2}$ are unaffected by the value of $D_-$, as they occur at very high densities where H$_-$ photodetachment is subdominant.

Quantitatively, the factor 3.5 shift in $n_{\rm cool, H_2}$ may be understood as follows.
For the higher $D_-$ rate, photodetachment dominates H$^-$ removal for densities $n < 10^2 I_{\rm UV} T_3^{0.39}$ cm$^{-3}$.
For such densities, $\eta \approx 0.1 T_3^{-0.39} I_{\rm UV}^{-1} (n/10{\rm cm^{-3}})$ (see Eq.~\ref{eq: eta}), and the H$_2$ abundance is 
\begin{equation}
\label{eq: xH2 gas high D-}
x_{\rm H_2} =  \frac{x_{\rm e} k_{\ref{reac: H- formation}} k_{\ref{eq: H2 form gas reac}} n^2}{D_0 D_- I_{\rm UV}^2} = 3.3 \times 10^{-8} T_3^{0.62} \zeta_{-16}^{1/2} I_{\rm UV}^{-2} \left( \frac{n}{10\ {\rm cm^{-3}}}\right)^{3/2}  \ .
\end{equation}
At $T_3=6$ and $n=10$ cm$^{-3}$ (i.e., near the Ly$\alpha$-to-H$_2$ transition) this $x_{\rm H_2}$ is a factor of $\approx 20$ smaller than the corresponding value in Eq.~(\ref{eq: xH2 gas}), and has a steeper dependence on density.
Plugging this back into Eq.~(\ref{eq: xH2 gas high D-}), we see that $n$ will now enter with a power 5/2, thus the factor 20 lower $x_{\rm H_2}$, needs to be compensated by an increase in density by a factor of $\approx 20^{2/5} \approx 3.3$, in agreement with our numerical results.
% 

% Plugging Eq.~(\ref{eq: xH2 gas high D-}) back into Eq.~(\ref{eq: point of LyA=H2, 0})
% we get
% \begin{equation}
% \label{eq: point of LyA=H2}
% n_{\rm cool,H_2} = \left( \sqrt{\frac{\alpha_B}{\zeta}} \frac{D_0 D_- I_{\rm UV}^2}{k_{\ref{reac: H- formation}}k_{\ref{eq: H2 form gas reac}}} \ \frac{E_{\rm cr} \zeta_p}{\mathcal{L}_{\rm H_2}} \right)^{2/5} 
% \approx 5.0 \ \zeta_{-16}^{4/5} I_{\rm UV}^{4/5} \ \ \ {\rm cm^{-3}} \ \ .
% \end{equation}
% 

Importantly, while $\zeta$, $I_{\rm UV}$ and $D_-$ affect the values of the critical densities and metallicities, the qualitative behaviour, the shift of the WNM-to-CNM transition to higher pressures  with decreasing metallicity, the quenching of the multiphase zone due to H$_2$ cooling at low $Z'$, and the final disappearance of the multiphase phenomenon at very low $Z'$, is robust.

\section{Summary and Discussion}
\label{sec: conclusions}

% We have studied the thermal phase structure of the neutral, predominantly atomic interstellar medium, across a wide range of metallicities, from supersolar to effectively pristine primordial gas, and for varying UV fields and cosmic-ray ionization rates.
% We include H$_2$ cooling and heating processes, which govern the phase properties at low metallicities. 
% The dominant processes are H$_2$ ro-vibrational cooling, and UV pumping and H$_2$ formation heating. 

In this paper we have studied the thermal phase properties of atomic gas from high to vanishing metallicity, with special emphasis on understanding how multiphase behavior is altered when H$_2$ heating and cooling processes become dominant. As discussed in \S \ref{sec: intro}, the critical role of H$_2$ cooling at low metallicity has long been recognized in many studies.  However, multiphased CNM/WNM behavior into the low metallicity and primordial regimes has received less attention.

For solar metallicity, we reproduce the known result that a CNM phase cooled by \cii and \oi line emission, ($T_{\rm CNM} \sim 100$ K) and a WNM phase cooled by Ly$\alpha$ emission ($T \sim 6000$ K) phases can coexist at thermal equilibrium at the same thermal pressure (\citealt{Field1969, Wolfire1995, Liszt2002}, \citetalias{Wolfire2003}).
% The multiphase is characterized by the maximum and minimum pressure in the pressure density curve, $P_{\rm min},P_{\rm max}$ and $n_{\rm min},n_{\rm max}$, which correspond to the minimum CNM and maximum WNM pressure and density.  
The WNM-to-CNM transition and the multiphase region shifts to higher densities and pressures with increasing 
UV intensity, $I_{\rm UV}$, and/or increasing ionization rate, $\zeta$, as photoelectric (PE) heating becomes more efficient.

As the metallicity decreases, the metal cooling rate decreases.
However, as long as  $Z' \gtrsim 0.1$, PE heating continues to dominate so that the heating rate also decreases. 
For sufficiently low metallicities, $Z_d'$ is expected to scale 
superlinearly
with $Z'$ (in our model this occurs for $Z'<0.2$), PE heating then falls faster than metal cooling, as $Z'$ decreases, and the CNM is then colder and is less dense compared to  solar metallicity models.
The pressure range that allows a multiphase is then larger  compared to solar metallicity models.

For metallicities $Z' \lesssim 0.1$, the ionization of neutral gas by cosmic-rays (CR) becomes the dominant heating mechanism of the gas (the metallicity value depends on the CR ionization rate).
In this limit as the metallicity decreases, the metal cooling rate decreases, while heating remains constant, shifting the WNM-to-CNM transition to higher densities and pressures.
In this limit, and when H$_2$ heating and cooling processes are ignored, we find that the Ly$\alpha$-to-metal transition density (and associated WNM-to-CNM transition) scales as $\zeta/Z'$.

However, while previous analytic models of the H{\small I} phase structure often ignored H$_2$ cooling, we find that  H$_2$ cooling dominates and modifies the phase structure at
low metallicity (as we show H$_2$ cooling and heating also play a role at solar metallicity gas but only at very high densities, $\sim 10^6$ cm$^{-3}$, deep in the CNM).
The critical metallicity below which H$_2$ cooling becomes important is $Z'_{\rm cool,H_2} = 8.0 \times 10^{-3} (\zeta_{-16}/I_{\rm UV})^{2/3}$.
Below this metallicity, H$_2$ cooling reduces the temperature of the WNM, to temperatures as low as 600 K (the exact value depending on metallicity).
This lowers $P_{\rm max}$ and results in a narrower multiphase zone, both in pressure and density.
% At metallicity of $Z' \approx 10^{-4}$, the multiphase region is already negligible (however, there still exist a moderate jump in tempearture from the WNM to the CNM, from $\approx 600$ to 200 K).

Finally, at extremely low metallicities $Z'<Z'_{\rm heat, H_2} = 1.2 \times 10^{-5} \zeta_{-16}^{2/3}$, the density at the WNM-to-CNM transition is already so high that H$_2$ heating starts contributing.
In this limit the thermal structure is determined only by H{\small I} (Ly$\alpha$) and H$_2$ cooling. 
There is a single-phase solution at any pressure, as the temperature decreases smoothly with gas pressure,  from the $\approx 10^4- 6000$ K  WNM, down to $T \approx 600$ K, as determined by the balance of H$_2$ cooling and heating.

Gravitational collapse is favored in the cold-dense CNM, where the free-fall time is relatively short.
Our result that the transition from warm to cold gas occurs at ever increasing pressures as the metallicity is reduced (over the range $10^{-5} \lesssim Z' \lesssim 0.1$, for a constant $I_{\rm UV}$ and $\zeta$), suggests that galaxies  with metal-poor ISM (i.e., galaxies in early evolutionary stages or low mass dwarf galaxies), should have maintained very high interstellar pressures in order to allow the formation of a CNM phase.
For example, for our fiducial model of $I_{\rm UV}=\zeta_{-16}=1$, and for a metallicity $Z'=10^{-3}$ a minimum pressure, $P/k \approx 10^5$ cm$^{-3}$ K is required.
This requires abnormally large ISM column densities (for hydrostatic pressure equilibrium), deep gravitational potential wells of the dark-matter halos, or a highly pressurized hot ambient gas 
% (e.g., as that discussed in \citet{McKee1977}, but with a higher density/temperature) 
that may provide the needed pressure support.

Alternatively, a WNM-to-CNM transition and a multiphase ISM may still occur at normal ISM pressures at low metallicity  if the heating rate is reduced.
This requires a reduced $\zeta$ and $I_{\rm UV}$ values, suggesting lower star-formation (supernovae) rates in low metallicity galaxies.
For example, for standard Galactic conditions ($Z'=I_{\rm UV}=\zeta_{-16}=1$), the multiphase occurs at $P\approx 3000$ cm$^{-3}$ K (Fig.~\ref{fig: T_2d}).
For  $Z'=10^{-3}$ the multiphase may still occur at a similar pressure if $I_{\rm UV}$ and $\zeta$ are reduced by a factor of 10 (Fig.~\ref{fig: T_2d_vars}, rightmost column, 3$^{\rm rd}$ row).
When connected to a theory of a self-regulated star forming ISM \citep{Ostriker2010}, this would imply that  the star-formation rate per unit gas column (i.e., the normalization of the Kennicutt Schmidt relation) should be lower at very low metallicities compared to solar (e.g., a factor of 10 lower for $Z'=10^{-3}$ compared to $Z'=1$).
If neither of the above mechanisms are sufficient to maintain the high pressure needed, the ISM would remain warm, at $\approx 10^4$ K.
In such a case, a different mode of star-formation would be expected, with only very massive clouds susceptible to gravitational collapse.
% This would significantly affect the star-formation efficiency and the initial mass function (IMF) of the young galaxies.

%At extremely low metallicities, below $Z'_{\rm h} =1.3 \times  10^{-4}$, we find that the multiphase phenomena is washed out.
Previous studies have suggested that the atomic ISM of star-forming galaxies is naturally driven towards the multiphase pressure range, which provides a regulation mechanism for star-formation  \citep{Parravano1988, Parravano1989, Ostriker2010}.
The idea is as follows.
If the star-formation rate of a galaxy increases this leads to an increase in the UV intensity which in turn increase the PE heating rate.
The $P-n$ curve is then pushed up and to the right  (see Fig.~\ref{fig: Tn_Z1}, red-dashed curve), the gas is driven to the warm (WNM) phase, resulting in a decrease in the efficiency of star-formation, and vice versa (see Fig.~1 in \citealt{Ostriker2010}).
In the limit where CR dominates over PE heating, the regulation is also expected to hold,
since CRs
are accelerated in supernovae remnants, and thus the CR ionization rate is also likely proportional 
 to the star-formation rate.
 
 If multiphase structure is indeed fundamental for star-formation regulation in galaxies, then our model provides a prediction for the thermal  
 pressure of the atomic ISM in galaxies, as a function of metallicity, CR ionization rate and UV intensity (i.e., see Figs.~\ref{fig: T_2d}, \ref{fig: T_2d_vars}).
 This pressure may be then related to the star-formation rate and galaxy gas and dark matter halo, through the requirement of hydrostatic equilibrium.

At extremely low metallicities, the multiphase structure disappears, and the multiphase-regulation mechanism cannot hold.
This may imply a different (potentially a bursty) mode of star-formation for galaxies in this early evolutionary stage.
In this limit the dense gas remains warm, at $\approx 600$ K, and is cooled by H$_2$ cooling. This temperature is a factor of $\sim 10$ higher than the classic CNM at solar metallicity.
The corresponding \citet{Jeans1902} mass is a factor of $\approx 10^{3/2} \approx 30$ higher.
Stellar structures forming in this low metallicity regime are expected to be more massive compared to their high metallicity counterparts.

Our finding that the cooling of the atomic ISM at low metallicity (below $\approx 0.01$) is often dominated by H$_2$ line emission from the moderately warm phase, may open up a new observational window for the study of the low metallicity ISM, in local dwarfs and in galaxies of early evolutionary stages at high redshift.
This is particularly timely with the upcoming launch of the James Webb Space Telescope (JWST) which will be able to observe IR H$_2$ lines at extreme sensitivity.

%AMIEL77 more adjustements
\acknowledgements
We thank John Black, Emeric Bron, Franck Le Pettit, Evelyne Roueff, Chris McKee, and Eve Ostriker,
for fruitful discussion and helpful suggestions.
 We thank the referee for constructive comments that improved our manuscript significantly.
SB thanks the Center for Computational Astrophysics at the Flatiron Institute for hospitality and funding where some of this research was carried out with AS.
This work was also supported by the German Science Foundation via DFG/DIP grant STE 1869/2-1 GE 625/17-1 at Tel Aviv University.

\appendix
\section{A.~Heating and Cooling Processes}

%shmuel6: Section modified
\subsection{Line-Emission Cooling}
In general, gas cooling through line-emission occurs whenever a particle (ion/atom/molecule) is collisionaly excited to a higher energy state, and then relaxes back to a lower energy state by (spontaneously) emitting a photon.
In this process, part of the kinetic energy of the collider particle is transferred to the emitted photon, leading to gas cooling.

While we compute the cooling rates numerically (as outlined below), it is instructive to consider the simple case of a two-level system (e.g., the \cii 158 $\mu$m transition), for which an analytic expression for the cooling rate may be obtained.
Let $\Delta E$ be the transition energy, and $x_i$ and $x_{j}$ the abundances of the cooling species and the collisional partners, respectively. 
The cooling rate per unit volume (erg cm$^{-3}$ s$^{-1}$) is 
\begin{equation}
\label{eq: L_cool two levels}
L_i \ = \  \ x_i n^2 \ \overbrace{ \sum_j x_{j} \ q_{ij,{\uparrow}}(T)\Delta E}^{\mathcal{L}(T)} \ \frac{1}{1+n/n_{{\rm crit}}} \ ,
\end{equation}
where  $n_{{\rm crit}} \equiv A/q_{i,\downarrow}$ is the critical density at which the collisional deexcitation and radiative decay rates are equal, $A$ is the Einstein coefficient for spontaneous decay, and $q_{ij,{\uparrow}}$ and $q_{ij,{\downarrow}}$ the collisional rate coefficients for the upward and downward transitions respectively.
The latter are related through $(q_{ij,{\uparrow}}/q_{i,{\downarrow}})=(g_u/g_d) \exp[-\Delta E/(k_B T)]$, where $g_u$, $g_d$, are the quantum degeneracies.
The combination $\sum_j x_{j} q_{ij,{\uparrow}}(T) \Delta E \equiv \mathcal{L}_i(T)$ is the {\it cooling coefficient}. 
For densities $n \ll n_{{\rm crit}}$, the cooling rate $L \propto n^2$, as the density of the coolant and the rate of collisions (which populate the upper level) are both $\propto n$.
At $n \gg n_{{\rm crit}}$ the energy levels are thermalized and $L \propto n$. In this limit,
\begin{equation}
\label{eq: cooling high n}
L_i \simeq x_i n \ n_{{\rm crit}}(T) \ \mathcal{L}(T) \equiv x_i n \ \Lambda_{{\rm LTE}}(T)  \ ,
\end{equation}
where $\Lambda_{{\rm LTE}}$ is cooling rate per  atom/molecule (erg s$^{-1}$) at local thermal equilibrium (LTE).
Eq.~(\ref{eq: L_cool two levels}) is exact for two level systems (assuming optically thin gas), but it may also provide a useful approximation for multilevel systems (such as \oi and H$_2$), with the appropriate choice of an effective $\Delta E$ and $n_{\rm crit}$.
% We use this approximation to derive analytic expressions. 

% In our numerical calculations, we solve for the full multi-line systems for the various coolants.
We calculate the line emission cooling for Ly$\alpha$, \cii 158 $\mu$m, \oi 63 and 146 $\mu$m, and the ro-vibrational transitions of H$_2$  and HD.

{\bf Ly$\alpha$ cooling:} We compute the cooling via radiative decay from the 2p and 2s states of neutral hydrogen collisionaly excited by electrons. 
For the collisional de-excitation rate coefficient, we adopt $q_{\downarrow} = 8.63\times 10^{-6} \Omega/(g_u \sqrt{T/{\rm K}})$ cm$^3$ s$^{-1}$, $g_u = 2 (2 l_u +1)$ where $l_u=0$ and $1$ for 2s and 2p is the quantum angular momentum number, and $\Omega$ is the dimensionless collision strength \citep{Gould1970, Spitzer1978a}.
We adopt a constant $\Omega = 0.26$ for the 2s level, and $0.4$ for 2p. 
The resulting total (2s+2p) collisional de-excitation rate coefficient is $q_{\downarrow}=1.7 \times 10^{-8}/\sqrt{T_4}$ cm$^3$ s$^{-1}$,
in excellent agreement with \citep{Gould1970, Spitzer1978a} as well as with the more recent calculations of \citet{Callaway1987}, over the temperature range of interest, $6000<T<10^4$ K.

{\bf Metal fine-structure cooling:} 
We calculate the cooling arising from the from the fine-structure transition of \cii $^2P_{3/2}^{\rm o}-^2P_{1/2}^{\rm o}$ at $\lambda=157.7$ $\mu$m, and the fine structure transitions of \oi $^3P_i-^3P_j$  where $[i,j]=([2,1],[2,0],[1,0])$, with the most important being the $[i,j]=[1,0]$ at $\lambda=63.2$ $\mu$m.
We include collisional excitations with electron, hydrogen atoms and para-H$_2$ and ortho-H$_2$.
We use the analytic fits presented in \citet[][see his Table 6]{Draine2011}, which are based on \citet{Pequignot1996, %e-O
Tayal2008, %e-CII
Barinovs2005,%CII-H
Flower1977, %CII-H2
Abrahamsson2007, %O-H
Jaquet1992}. %O-H2
We also include cooling from [C{\small I}] (although we find it to be a subdominant coolant), using the analytic fits from  \citet[][see his Table 6]{Draine2011}, based on data from \citet{Roueff1990, Abrahamsson2007, Schroder1991, Staemmler1991}.

{\bf H$_2$ ro-vibrational cooling:} 
For the H$_2$ cooling function we use the polynomial fits provided by \citet{Glover2008} with updates from \citet[][see their Appendix A1.2]{Glover2015b}, which account for excitations of the H$_2$ ro-vibrational ladder through collisions with 
 H, He, para-H$_2$, ortho-H$_2$, H$^+$, and e, assuming a constant ortho-to-para ratio o/p=3.
 We find that our results are not sensitive to the adopted o/p value.
The cooling functions are based on collisional data from \citet{Ehrhardt1968, Crompton1969, Linder1971,  Draine1983, Gerlich1990, Krstic2002, Wrathmall2007, Flower1998, Flower1998a, Balakrishnan1999, Flower1999, Honvault2011}. 
We have also calculated the H$_2$ cooling efficiency assuming the updated H-H$_2$ collision rates of  \citep{Lique2015}. We find that adopting this alternative collisional dataset results in at most a negligible difference in the total H$_2$ cooling function (which includes all the collisional partners).

{\bf HD ro-vibrational cooling:} 
For HD cooling we use the cooling function of \citet{Lipovka2005} 
\footnote{We spotted typos in the formulae in \citet{Lipovka2005} (confirmed by the authors): (a) $T$ and $n$ in their Eq.~(4) and (5), should be replaced with $\log T$ and $\log n$. (b) In their Eq.~(5), $W_{\rm HD}$ should be replaced with $W_{\rm HD}/n$.}, based on collisional data from \citet{Roueff1999, Roueff1999a}.

% based on H2-H collisional data (Tine, Lepp Dalgarno 1998)

\subsection{Heating and Cooling by Dust} 
\label{sub: app_heat_cool}
Dust grains and PAHs, contribute both to gas heating, through the injection of energetic photoelectrons, and to cooling, though dust/PAH-assisted recombination.
For the photoelectric (PE) heating 
we follow \citet{Bakes1994} and \citetalias{Wolfire2003} and adopt
\begin{equation}
\label{eq: p.e. heating}
G_{\rm pe} = \gamma_{\rm pe} \epsilon Z_d' I_{\rm UV} n \ ,
\end{equation}
where the PE efficiency is
\begin{equation}
\label{eq: p.e. epsilon}
\epsilon = \frac{4.9 \times 10^{-2}}{1+5.9 \times 10^{-13} y^{0.73}} + \frac{3.7\times 10^{-2} T_4^{0.7}}{1+3.4\times 10^{-4} y}
\end{equation}
and
\begin{equation}
\label{eq: p.e. y}
    y \equiv \frac{I_{\rm UV} (T/{\rm K})^{1/2}}{(n_e/{\rm cm^{-3}}) \phi_{\rm PAH}} \ .
\end{equation}
In Eq.~\ref{eq: p.e. heating} and \ref{eq: p.e. epsilon}, 
we assume that the PAH abundance scales with the dust-to-gas ratio, $Z_d'$, and adopt
$\gamma = 2.2\times 10^{-24}$ erg s$^{-1}$
and $\phi_{\rm PAH}=0.5$ (see Eqs.~19-21 in \citetalias{Wolfire2003}
\footnote{The numerical values in Eq.~\ref{eq: p.e. heating} and \ref{eq: p.e. epsilon} differ from those presented in \citetalias{Wolfire2003} because we used $I_{\rm UV}$ as the basic UV intensity unit, where \citetalias{Wolfire2003} use the \citet{Habing1968} $G_0$ units. The conversion is, $1 I_{\rm UV}=1.7 G_0$.}).
For recombination cooling we adopt
\begin{equation}
    L_{\rm rec} = \gamma_{\rm rec} (T/{\rm K})^{0.94} (1.7 y)^{\beta} \phi_{\rm PAH} \ Z_d' \ x_{\rm e} \ n^2 \ ,
\end{equation} 
with $\gamma_{\rm rec}=4.65 \times 10^{-30}$ erg cm$^3$ s$^{-1}$.

%shmuel6: modified
In our model, we do not solve for the PAH chemistry. 
At high metallicity, the PAHs may affect the electron fraction as they introduce an additional recombination channel for the H$^+$ (see \citetalias{Wolfire2003}).
At low metallicity, this effect becomes unimportant as the abundance of the PAHs falls rapidly with the vanishing DGR.
Even at high metallicity, where the PAHs are importnat for recombination, we find that the net effect on the phase diagrams is mild, as shown in Fig.~\ref{fig: T_2d_vars} where we compare to \citetalias{Wolfire2003} (see also Fig.~8 in \citetalias{Wolfire2003} for the effect of PAH abundance on the phase diagrams).
% Within the parameter space we consider dust-gas collisional heating/cooling  always plays a subdominant role.

We include cooling due to collisions of gas particles with dust-grains. 
We adopt 
\begin{equation}
    L_{\rm g-d} = \gamma_{\rm g-d}  (T/{\rm K})^{1/2}  (T - T_d)  \left(1 - 0.8 \mathrm{e}^{-75 {\rm K}/T} \right) Z_d' n^2 \ ,
\end{equation}
where $T_d$ is the dust temperature, and
$\gamma_{\rm g-d}=3.8 \times 10^{-33}$ erg cm$^3$ s$^{-1}$, \citep{Hollenbach1989, Glover2007b}.
The dust is typically much colder than the gas, and is approximated by $T_d=16.4 I_{\rm UV}^{1/6}$ K \citep{Draine2011}.

\subsection{Cosmic-Ray Heating}
Cosmic-ray or X-ray ionization liberate energetic electrons into the gas. These energetic electrons lose energy through coulomb interactions, and ionization, dissociation and excitations.
The fraction of energy that is lost into coulomb interactions heat the gas.
We adopt
\begin{equation}
\label{eq: CR heating}
G_{\rm cr} = \zeta_{\rm p} E_{\rm cr} n \ ,
\end{equation}
for the CR heating rate, where $\zeta_{\rm p}=\zeta/(1+\phi_s)$ is the primary CR ionization rate,
\begin{equation}
\label{eq: phi_s}
\phi_s = \left( 1 - \frac{x_{\rm e}}{1.2} \right) \frac{0.67}{1+(x_{\rm e}/0.05)} \ \ \ ,
\end{equation}
is the number of secondary ionizations per each primary electron, and
\begin{equation}
E_{\rm cr} \ = \ 6.43 \left(1+ 4.06\left[\frac{x_{\rm e}}{x_{\rm e}+0.07}\right]^{1/2} \right) \ {\rm eV} \ ,
\end{equation} is the characteristic deposition energy per ionization event \citep{Dalgarno1972, Draine2011}. 
The free electrons in the gas can also interact directly with the CR protons, giving an additional heating rate
\begin{equation}
G_{\rm cr}= \zeta_{\rm p} E_{\rm cr} n x_{\rm e} \ .
\end{equation}
For CR proton energies of $\sim 50$ MeV, $E_{\rm cr}=287$ eV  \citep{Goldsmith1969}.
This direct heating channel dominates when the electron fraction is larger than $x_{\rm e} \sim 0.02$.

\subsection{H$_2$ Heating} 
%shmuel6: SubSection modified

When LW photons are absorbed by H$_2$ molecules, they lead to population of the  excited electronic (LW) states.
$\sim 10$~\% of the radiative decays lead to H$_2$ photodissociation, whereas in the rest $90$~\% of the cases, the molecule decays through ro-vibrational transitions back to the ground state.
If the gas density is sufficiently high, such that collisions rather than spontaneous emission deexcite the H$_2$, the energy of the excited level is transferred to the colliding particle (H, e, H$_2$, etc.) leading to gas heating.
Following \citet{Burton1990} and \citet{Rollig2006}, the pumping heating rate may be approximated with 
\begin{equation}
\label{eq: pump heating}
G_{\rm H_2, pump} =  \ 9 D_0 I_{\rm UV}  \ E_{\rm pump}  \ x_{\rm H_2} n \ \frac{1}{1+n_{\rm crit}/n} 
= \ 9  \ Rn^2 \  x_{\rm H} \ E_{\rm pump} \ \frac{1}{1+n_{\rm crit}/n} \ ,
\end{equation}
where $9 D_0$ is the H$_2$ pumping rate (9 excitations occur per dissociation), and $E_{\rm pump}$ and $n_{\rm crit}$ are the effective energy and critical density of the pseudo two-level transition (see discussion above).
The second equality holds for chemical equilibrium, for which $D_0 I_{\rm UV} x_{\rm H_2} = R n  x_{\rm H}$ where $R$ is the total H$_2$ formation rate coefficient.  
In this limit, the UV pumping heating rate is proportional to the H$_2$ formation rate, $R n^2$.
At low densities, below $n_{\rm crit}$, another factor of $n/n_{\rm crit}$ enters, reflecting the fact that when $n< n_{\rm crit}$ spontaneous emission dominates deexcitation, not collisions.
Following \citet{Rollig2006} we adopt $E_{\rm pump} = 1.12$~eV, and $n_{\rm crit} = 1.1 \times 10^5 / \sqrt{T_3}$ cm$^{-3}$ for the effective energy released per UV pumping event and for the critical density.

We also include heating through H$_2$ photodissociation, and adopt
\begin{equation}
\label{eq: pd heating}
    G_{\rm H_2, pd} =  D_0 \ I_{\rm UV}  \ E_{\rm pd} \ x_{\rm H_2} n \ = \ R \ n^2 \  x_{\rm H} \ E_{\rm pd}
\end{equation}
where $E_{\rm pd}=0.4$ eV \citep{Black1977}. 
Photodissociation heating dominates over pumping heating when $n/n_{\rm crit} < E_{\rm pd}/(9E_{\rm pump})$, which occurs at
$n \lesssim 4.4 \times 10^{3}/\sqrt{T_3}$ cm$^{-3}$, for our adopted parameters.

When a new H$_2$ molecule is formed, part of the binding energy (4.5 eV) is converted into heat, either by providing translational energy of the H$_2$ molecules, or by exciting the H$_2$ ro-vibrational  levels which when followed by collisional de-excitation lead to gas heating.
The latter is effective when the gas volume density is large, of order of or higher than the critical density of H$_2$.
The heating resulting from H$_2$ may be written as
\begin{equation}
\label{eq: form heating}
G_{\rm H_2, form} \ = \  x_{\rm H} \ n^2\left\{ \sum_i R_i \left( E_{i,{\rm form,1}} + E_{i,{\rm form,2}} \frac{1}{1+n_{\rm crit}/n}\right) \right\}
\end{equation}
where the summation is over the formation channels of H$_2$, (i) formation on dust grains, and (ii) the gas-phase H$^-$ formation sequence, and $R_i$ are given by Equations (\ref{eq: RZ}) and (\ref{eq: R gas phase}), for the dust and gas formation routes, respectively.
For each of the formation processes, the first term in brackets, $E_{{\rm form,1}}$, is the energy that goes into translational energy of the products (for dust-formation, this is the H$_2$. For the gas phase H$^-$ formation route, most of the energy goes to the electron), and  $E_{{\rm form,2}}$ is the energy that goes into internal molecule excitation, which can then heat the gas through collisional deexcitation.
We adopt $E_{\rm form,1}=0.2$~eV and $E_{\rm form,2}=4.48$~eV for the dust catalysis \citep[][see their Appendix VI(c)]{Hollenbach1979}.
For the H$^-$ formation we follow \citet{Cizek1998} who finds that on average $\approx 0.6$ eV is released as translational energy, and adopt $E_{\rm form,1}=0.6$ eV and $E_{\rm form,2}=3.13$ eV.

We may obtain a simple form for the total heating rate by all of the H$_2$ heating processes discussed above.
Assuming that always a single formation route dominates (either the dust catalysis or the gas phase H$^-$ route) and that the H$_2$ system has reached chemical equilibrium, we get 
\begin{equation}
\label{eq: H2 heating}
G_{\rm H_2} = R n^2 \ x_{\rm H} \left( E_{\rm H_2, 1} +  E_{\rm H_2, 2} \frac{1}{1+n_{\rm crit}/n}  \right) \  \ .
\end{equation}
Here we defined $E_{\rm H_2,1} = E_{\rm form, 1} + E_{\rm pd}$ and $E_{\rm H_2,2} = E_{\rm form, 2} + 9E_{\rm pump}$ and assumed that the effective critical densities are the same for H$_2$ formation and UV pumping. In practice, since these critical densities are effective two-level system approximations, their values may differ for different processes.
With our assumed parameters, $E_{\rm H_2, 1}=(0.6,1.0)$ eV and $E_{\rm H_2,2}=(14.6,13.2)$ eV for H$_2$ formation on dust and via H$^-$, respectively.
The second term in Eq.~(\ref{eq: H2 heating}) is dominated by H$_2$ pumping heating, and becomes greater than the first term when $n \gtrsim (5.0, \ 8.4) \times 10^3/\sqrt{T_3}$ cm$^{-3}$, for H$_2$ formation on dust and via H$^-$, respectively.

\section{B.~Reduced Chemical Network}

To improve computational speed, we consider a reduced chemical network consisting of the species: H, H$^-$, H$^+$, H$_2$, He, C$^+$, O, and e. 
The chemical reactions considered in the reduced network are listed in Table \ref{table: reduced netwrok}, along with the rate coefficients (based on the UMIST 2012 database, \citealt{McElroy2013}).
 For a thorough discussion of these chemical reactions see \S \ref{sub: params: R_form}. In summary,
\begin{table*}[h!]
\caption{Reduced Network}
\centering % used for centering table
\begin{tabular}{l l l}
\hline\hline %inserts double horizontal lines
Reaction & Rate coefficient (subscript refers to the reaction no.~in the text) & \\ [0.5ex]
   \hline 
${\rm H \ + \ H:dust \ \rightarrow \ H_2 }$ & $R_d=3\times 10^{-17} T_2^{1/2} Z_d'$ cm$^3$ s$^{-1}$\\
${\rm H \ + \ e \ \rightarrow \ H^- \ + \ \nu}$ & $k_{\ref{reac: H- formation}} = 7.2 \times 10^{-16} T_3^{0.64} \ {\rm e}^{-9.2 {\rm K}/T}$ cm$^3$ s$^{-1}$ \\
${\rm H^{-} \ + \ H \ \rightarrow \ H_2 \ + \ e }$ & $k_{\ref{eq: H2 form gas reac}}=2.6 \times 10^{-9} T_3^{-0.39} {\rm e}^{-39.4 {\rm K}/T}$ cm$^3$ s$^{-1}$\\
${\rm H^- \ + H^+ \ \rightarrow \ H \ + \ H}$ & $k_{\ref{reac: mutual neutralization}}=4.1\times 10^{-8} T_3^{-0.5}$ cm$^3$ s$^{-1}$\\
${\rm e \ + \ H^+ \ \rightarrow \ H \ + \ \nu}$ & $\alpha_B = 1.4\times 10^{-12}T_3^{-0.75}$ cm$^3$ s$^{-1}$\\
${\rm H_2 \ + \ H \ \rightarrow \ H \ + \ H + \ H}$ & Formula from \citet{Martin1996} \\
[0.5ex]
   \hline    \hline 
Reaction & Photo/CR-rate& \\ [0.5ex]
   \hline 
${\rm H \ + \ CR \ \rightarrow \ H^+ \ + \ e}$ & $\zeta = 10^{-16} \zeta_{-16}$ s$^{-1}$\\
${\rm H_2 \ + \ \nu \ \rightarrow \ H \ + \ H }$ &  $D_0 I_{\rm UV} = 5.8\times 10^{-11} I_{\rm UV}$ s$^{-1}$\\
${\rm H^- \ + \nu \ \rightarrow \ H \ + \ e}$ & $D_-I_{\rm UV} = 5.6\times 10^{-9} I_{\rm UV}$ s$^{-1}$\\
C$^+$, O, He & assume all gas-phase carbon/oxygen/helium are in C$^+$, O, He \\
\hline %inserts single line
%\caption*{}
\end{tabular}
\label{table: reduced netwrok} % is used to refer this table in the text
\end{table*}
the
H$_2$ is formed via dust catalysis and by the H$^-$ gas-phase route, and is destroyed by photodissociation.
The H$^-$ is formed via radiative association, and is destroyed by associative detachment,
photodetachment, and
mutual neutralization.
For the electrons we include production by CR-ionization of hydrogen, and photoionization of carbon assuming all the carbon is in the form of C$^+$. 
The electrons are removed by radiative recombinations the protons. We assume
the oxygen and helium are in neutral atomic form. The proton abundance is set by the requirement of charge neutrality.

This reduced network enables us to obtain analytic expressions for the species abundances and thus to improve significantly the computation speed. 
The reduced network was used to produce Figs.~\ref{fig: T_2d} and  \ref{fig: T_2d_vars}.
We have confirmed the accuracy of the reduced network by comparing the equilibrium $T(n)$, and $x_{\rm H_2}(n)$ curves for the reduced and full networks at the various metallicities considered in Fig.~\ref{fig: Tn_diff_Z}.

% With the reduced network we derive the following analytic equations for the species abundances.
% First we compute the electron abundance with
% \begin{equation}
% \label{eq: min_net_xe}
% x_{\rm e} = 
% \end{equation}
% This expression is valid for predominantly neutral gas, i.e., $x_{\rm e}<1$.
% In Eq.~(\ref{eq: min_net_xe}), $x_{\rm C^+}=A_{\rm C}Z'$, and the proton abundance is given by $x_{\rm H^+}=x_{\rm e}-x_{\rm C^+}$.
% The H$^-$ abundance is given by
% \begin{equation}

% \end{equation}

 \section{C.~Time-scales}
\label{sub: timescales}
%Shmuel5: I changed this entire section
Thermal and chemical equilibrium are good assumptions whenever the dynamical time of interest, $t_{\rm dyn}$, e.g., the turbulent crossing time, the free-fall time, the cloud lifetime, etc., is long compared to the cooling time, $t_{\rm cool}$, and the chemical time, $t_{\rm chem}$.

%As we now discuss the cooling time is typically short in the CNM, but becomes long, of order of 10 Myrs, in the WNM, especially at low metallicity. 
%The chemical time is typically short, shorter 

% as it is set by the H$_2$ photodissociation rate, which is typically very rapid.
\subsubsection{The cooling/heating time}
The cooling time from an initial temperature $T_i$ to an equilibrium temperature $T$ is
\begin{equation}
    t_{\rm cool} = \int_{T_i}^{T} \frac{(3/2)  n k_B T}{G(n,T)-L(n,T)} \mathrm{d}T \ ,
\end{equation}
where we assumed isochoric cooling of predominantly atomic gas (an adiabatic index $\gamma=5/3$).
If the cooling time is dominated by the cooling rate near the final equilibrium temperature, the cooling time simplifies to
\begin{equation}
\label{eq: t_cool}
    t_{\rm cool} \approx  \ \frac{(3/2)  n k_B T}{L(n,T)} \approx \frac{(3/2)  n k_B T}{G(n,T)} \ 
    .
\end{equation}
This is a good assumption when the cooling function increases superlinearly with  $T$, as occurs when H$_2$ or Ly$\alpha$ emission dominate the cooling, for $T \gtrsim 6 \times 10^3$ K, and if $Z'<Z'_{\rm cool, H_2}$ down to $\sim 1000$ K (see discussion in \S \ref{sec: low Z}, and Fig.~\ref{fig: cooling functions} below), or at low temperatures, $T \lesssim 100$ K where metal cooling dominates but is a steep function of the temperature.
On the other hand, When metals dominate the cooling at $T \gtrsim 100$ K,  the cooling function depends weakly on $T$ (with a power-index $<1$), and the cooling time is then governed by the initial conditions. In this case the cooling time may still be approximated by
Eq.~(\ref{eq: t_cool}), but with $T$ replaced by $T_i$.

In the second equality in Eq.~(\ref{eq: t_cool}) we used the fact that the cooling time is dominated by the time spent near the equilibrium temperature, where $G \approx L$.
This latter form of $t_{\rm cool}$ is particularly useful when $G$ scales approximately linearly with the density while being weakly dependent on temperature.
For example, for CR heating, $G = \zeta_p E_{\rm cr} n$ and we get
\begin{equation}
\label{eq: t_cool_2}
    t_{\rm cool} =  \ \frac{(3/2) k_B T}{\zeta_p E_{\rm cr}}
    \approx 42.4 T_4 \zeta_{-16}^{-1} \ {\rm Myr} .
\end{equation}
In the numerical evaluation we used $E_{\rm cr} = 15.2$ eV and $\phi_s = 1.57$ (the same values as in Eq.~\ref{eq: point of LyA=OI low Z}).
We see that for typical $\zeta_{-16}=1$, the cooling time in the WNM is long, with $t_{\rm cool} \sim 20$ Myrs, for $T = 6000$ K.
For the CNM, the cooling time is considerably shorter, with $t_{\rm cool}$ typically 0.1-0.4 Myrs Since the heating rate is mostly independent of temperature (e.g., for CR heating), 
Eqs.~(\ref{eq: t_cool}-\ref{eq: t_cool_2}) also express the heating time from $T_i$ to $T$ for any arbitrary $T_i$ and $T$.

%
%In the upper-right panel of Fig.~\ref{fig: T_2d} we
%show contours of the cooling times in the $Z'-n$ parameter space superposed on the color-coded equilibrium temperatures.
% The cooling times range from $\sim 1-10$ Myrs in the WNM, depending on the metallicity, 
%down to 0.1-0.4 Myrs in the CNM at low $Z'$, and down to less then 0.01  Myrs at high $Z'$.
%At higher metallicities the increase in the heating rate due to the onset of PE heating, is accompanied by an appropriately higher cooling rates. 
%For the CNM, the cooling rate naturally increase with increasing metallicity, as \cii and \oi dominates the cooling.
%In the WNM, where Ly$\alpha$ dominates, the cooling rate increases via a slight increase in the equilibrium temperature. 
%Since at low $Z'$ a large portion of the parameter space is WNM, or intermediate $\approx 1000$ K medium (at very low $Z'$), with cooling times 1-10 Myrs, the gas may potentially be out of thermal equilibrium, depending on the magnitude of the local dynamical time (e.g., the sound crossing time, turbulent time, free-fall time, etc.).

\subsubsection{Chemical Times}

Generally, the time for the H$_2$ (and H) abundance to reach steady-state is
\begin{equation}
    t_{\rm H_2} = \frac{1}{D I_{\rm UV} + 2Rn} \ .
\end{equation}
where $D$ is the local photodissociation rate.
In highly shielded cloud interiors, $D \ll 2 Rn$, and the H$_2$ time equal the H$_2$ formation time, $1/(2Rn)$, which may become long.
However, in our models we consider the state of the ambient atomic medium exposed to the free-space UV field, with $D=D_0$. 
In this regime the H$_2$ time is the H$_2$ (free-space) photodissociation time, $1/D_0=550 \ I_{\rm UV}^{-1} \ {\rm yr}$, and is always very short.

% At low metallicities, dust attenuation and H$_2$ self-shielding are both inefficient, and the dissociation rate $D \approx D_0$.
% In this regime, the H$_2$ time equals the (free-space) photodissociation time
%  $t_{\rm H_2} = 1/(D_0 I_{\rm UV}) = 550 \ I_{\rm UV}^{-1} \ {\rm yr}$, and is very short.

% At high metallicity and in regions of high volume density, self-shielding becomes significant and $D \ll D_0$, lengthening the H$_2$ time-scale. 
% When self-shielding becomes very effective, $D<2Rn$, the gas becomes predominantly molecular, and the H$_2$ time is the H$_2$ formation time, $t_{\rm H_2}=1/(2Rn)$.
% Assuming $R=3 \times 10^{-17}$ cm$^3$ s$^{-1}$, we get $t_{\rm H_2} \approx 5.3 (n_{2} Z_d')^{-1}$ Myrs.
% In this regime, the H$_2$ time is long compared to the cooling time, and 
% time-dependent effects may play important role.
% For example, if the gas is initially fully atomic, the  H$_2$ pumping and H$_2$ formation heating rates will be elevated relative to their equilibrium values
% \footnote{When $x_{\rm H_2}$ is small, process (a) becomes efficient as UV shielding is ineffective and thus the UV pumping rate is high. Process (b) is efficient because the H abundance is high elevating the H$_2$ formation rate.}
% , and as long as $t \ll t_{\rm H_2}$, the gas will remain warm (up to $T \sim 10^3$ K).
% Only once the H and H$_2$ have approached their equilibrium values, at $t \sim t_{\rm H_2}$,
% will the temperature fall to its equilibrium value, $T \lesssim 100$ K.

Other relevant chemical timescales are the e, C$^+$, and O, timescales.
% The electron abundance is most important fo Ly$\alpha$ cooling, and also plays a (minor) role in setting the metal and H$_2$ cooling efficiencies (see Fig.~\ref{fig: cooling functions}). 
At low  metallicity, the steady-state electron abundance is set by the ionization-recombination  equilibrium, as given by Eq.~(\ref{eq: xe}).
Under the assumption of $x_{\rm e} < 1$, the time-scale for this process to achieve steady-state is 
\begin{equation}
   t_{\rm e} = \frac{1}{\alpha_B n} = 0.13 T_4^{0.75} n_0^{-1} \ {\rm Myr} \ ,
\end{equation}
where $n_0 \equiv n/(10^0 {\rm cm^{-3}})$. The electron time is already short ($\ll 1$ Myr) in the WNM, and becomes extremely short in the CNM, with $t_{\rm e} \approx 40 T_2^{0.75}/n_2$ years.
For the C$^+$ time, consider a gas that recombines from a higher ionization state, ${\rm C^{+2} + e \rightarrow \rm  C^+}$. The recombination time is $t_{\rm C^+} = 1/(n x_{\rm e} \alpha_{\rm C^+}) \approx 4 (n_0 \zeta_{-16})^{-1/2}$  Myr, where we used $\alpha_{\rm C^+}=6.5 \times 10^{-12}$ cm s$^{-1}$ (\citealt{Nussbaumer1969}, see Table 14.7 in \citealt{Draine2011}),
and
used the equilibrium electron abundance (Eq.~\ref{eq: xe}; we can assume equilibrium for $x_{\rm e}$ since $t_{\rm e}\ll t_{\rm C^+}$).
The timescale for the chemical equilibrium of O may be longer than that of C$^+$, as the recombination rate for O$^+$  is longer than for C$^{+2}$. Nevertheless,  as long as C$^+$ has reached equilibrium, the effect on the temperature will not be significant, since both coolants operate at the same tempearture regime.

\end{document}